\RequirePackage[l2tabu, orthodox]{nag}
\RequirePackage{fix-cm}
\documentclass[11pt,onecolumn,letter]{IEEEtran}

 \usepackage{etoolbox}
\patchcmd{\section}{\centering}{}{}{}

\usepackage[a4paper, left=3cm, right=3cm, top=3cm, bottom=3cm]{geometry}

\usepackage{graphicx} 

\usepackage[cmex10]{amsmath} 
\usepackage{amssymb}
\usepackage{amsfonts}
\usepackage{mathrsfs} 

\usepackage{cuted}

\usepackage{subcaption}

\usepackage[noadjust]{cite}
\usepackage{epstopdf}

\usepackage[x11names]{xcolor}   

\usepackage[disable,colorinlistoftodos]{todonotes}  

\usepackage[T1]{fontenc}
\usepackage{charter}
\usepackage{environ}
\usepackage{tikz}
\usetikzlibrary{calc,matrix}

\usepackage[section]{placeins} 
\usepackage{stackengine}

\usepackage{enumitem}
\setlist[enumerate]{label=\roman*} 

\usepackage{microtype}

\newtheorem{theorem}{Theorem}
\newtheorem{corollary}{Corollary}
\newtheorem{conjecture}{Conjecture}

\usepackage[colorlinks]{hyperref}
\hypersetup{%
	colorlinks = true,
	citecolor=Green4,
	linkcolor  = blue
}

\pagebreak

\ifCLASSINFOpdf
\else
\fi

\begin{document}
\IEEEoverridecommandlockouts

\title{Heuristic Sensing Schemes for Four-Target Detection in Time-Constrained Vector Poisson and Gaussian Channels}

	\author{\IEEEauthorblockN{Muhammad Fahad\IEEEauthorrefmark{1},~\IEEEmembership{Member,~IEEE}, and Daniel R. Fuhrmann\IEEEauthorrefmark{2},~\IEEEmembership{Fellow,~IEEE}} \\
		\IEEEauthorblockA{Department of Applied Computing,\\
			Michigan Technological University\\ 
			Houghton, MI 49931, USA \\
			\IEEEauthorrefmark{1}mfahad@mtu.edu,
			\IEEEauthorrefmark{2}fuhrmann@mtu.edu}}

\maketitle
\begin{abstract}
In this work we investigate the different sensing schemes for detection of four targets as observed through a vector Poisson and Gaussian channels when the sensing time resource is limited and the source signals can be
observed through a variety of sum combinations during that fixed time. For this purpose we can maximize the mutual information or the
detection probability with respect to the time allocated to different sum combinations, for a given total fixed time. It is observed that for both Poisson and Gaussian channels; mutual information and Bayes risk with $0-1$ cost are not necessarily consistent with each other. Concavity of mutual information between input and output, for certain sensing schemes, in Poisson channel and Gaussian channel is shown to be concave w.r.t given times as linear time constraint is imposed. No optimal sensing scheme for any of the two channels is investigated in this work. 
\end{abstract}

\section{Introduction}  \label{intro}
In \cite{paper1}, \cite{paper2} and \cite{phdfahad} a two-target detection in vector Poisson and Gaussian channels is considered. It was observed that prior probability of the targets have a direct influence in deciding which of the sensing method is better over the other besides the given available sensing time. As, we study the problem in higher dimensions (or when targets are greater than two), we are hampered by the limitations of the  deterministic computational methods which fails to work efficiently, in terms of computational time, due to the curse of the dimensionality issue. Therefore, to study the problem in higher dimensions; resorting to some statistical computational method is one way to circumvent the exponentially rising dimensionality in the objective functions and Monte-Carlo method is used in this work \cite{hero2007foundations}, \cite{verdu2019empirical} and \cite{verdu2021error}.   

\begin{figure}[t]
	\centering
	\includegraphics[width=0.55\linewidth]{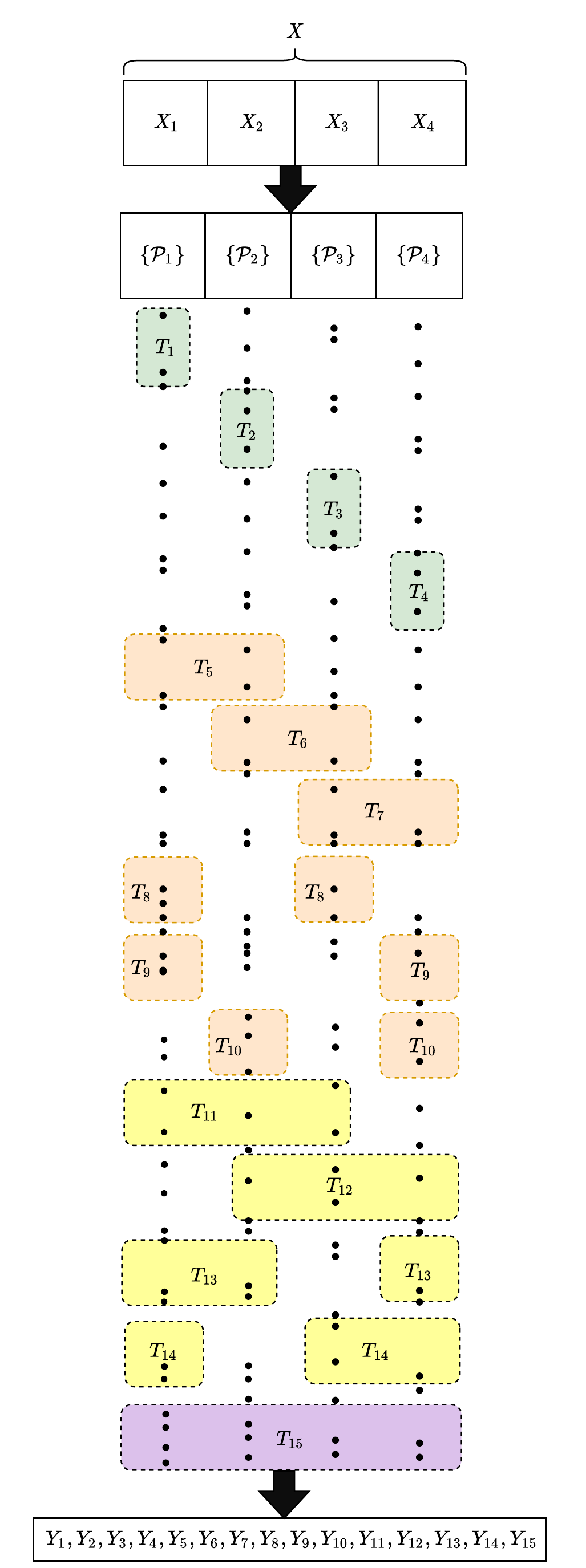}
	\caption{Illustration of sensing paradigm for detection of $4-$long hidden random vector $X$ from $15-$long observable random vector $Y$ through a vector Poisson channel under a total time constraint of $T=\sum_{i=1}^{15} T_i.$  \todo[inline]{1) Open draw.io in chrome. \newline 2) Open file from C:$\backslash$Users$\backslash$mfahad$\backslash$Dropbox$\backslash$Apps$\backslash$drawio$\backslash$Sensing\_Paradigm4.html \newline 3) Export to HTML and save in the same folder.  \newline 4) Open HTML and print PDF \newline 5) Crop the pdf using acrobats tools: crop  \newline 6) Save the pdf in folder C:$\backslash$Users$\backslash$mfahad$\backslash$Dropbox$\backslash$PhD\_Paper\_1 }}
	\label{f}
\end{figure}

\begin{figure}[t]
		\centering
		\includegraphics[width=0.9\linewidth]{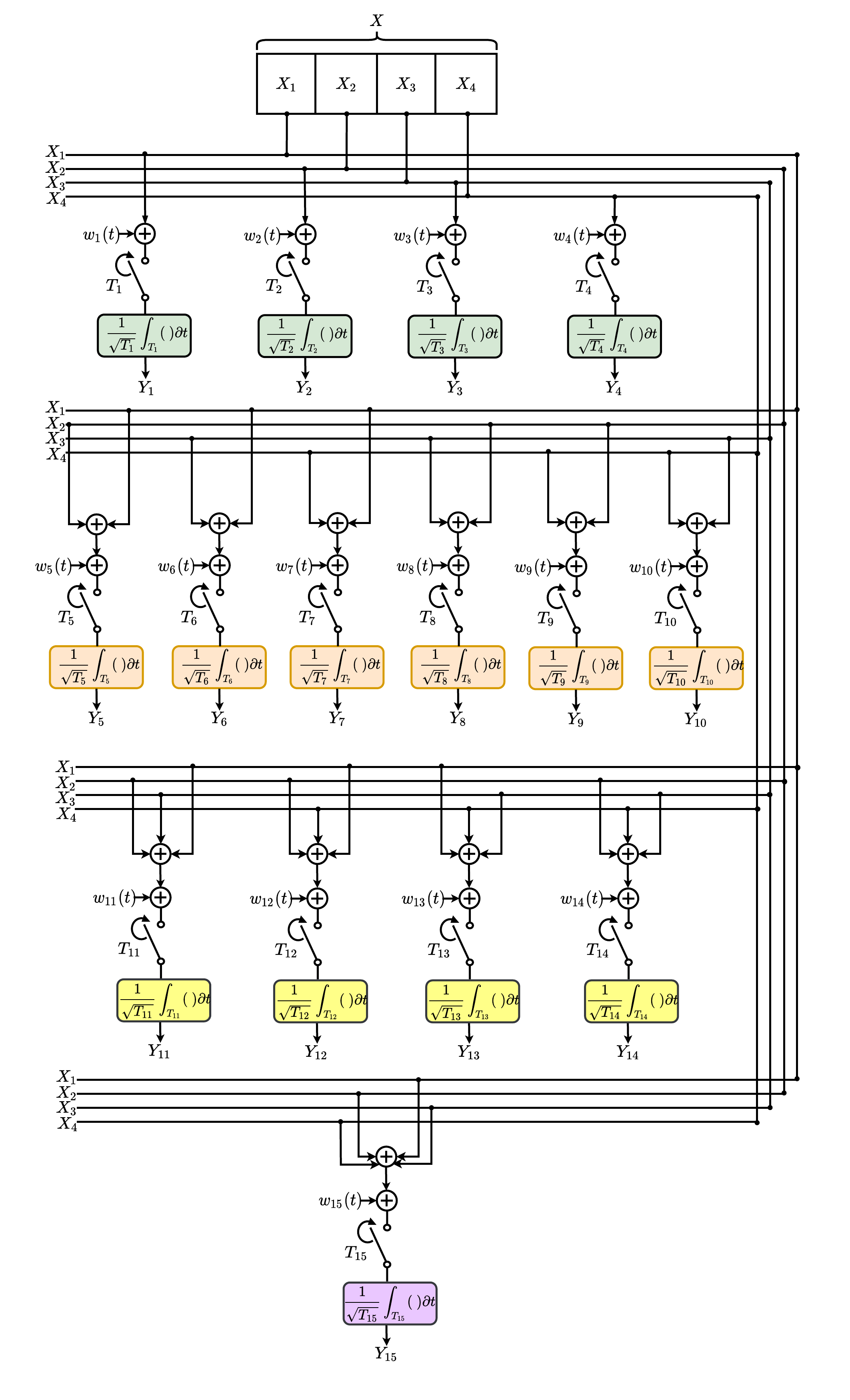}
		\caption{Illustration of sensing paradigm for detection of $4-$long hidden random vector $X$ from $15-$long observable random vector $Y$ through a vector Gaussian channel under a total time constraint. Where $w_i(t)$ are independent  white noise processes. Only one of the integrators becomes active for a time $T_i$ such that total time constraint $T=\sum_{i=1}^{15} T_i$ is satisfied by all the integrators. \todo[inline]{1) Open draw.io in chrome. \newline 2) Open file from C:$\backslash$Users$\backslash$mfahad$\backslash$Dropbox(4)$\backslash$Apps$\backslash$drawio$\backslash$Sensing\_Paradigm14.html \newline 3) Export to HTML and save in the same folder.  \newline 4) Open HTML and print PDF \newline 5) Crop the pdf using acrobats tools: crop  \newline 6) Save the pdf in folder C:$\backslash$Users$\backslash$mfahad$\backslash$Dropbox$\backslash$PhD\_Paper\_1 }}
		\label{f}
	\end{figure}

This paper considers an experimental design problem of setting, sub-optimally, the time-prop\-ortions for identifying a four-long binary random vector that is passed through a vector Poisson and vector Gaussian channels, and then based on the observation vector; classification of the input vector is performed and performance of any sensing scheme is then compared. Since, finding the optimal solution for the problem requires computations to be performed in (15-dimensional search-space as closed-form solutions don't exist); we have instead restricted to a reduced dimensional search-space and studied some sensing techniques that are sub-optimal.

We start by explaining the problem in the vector Poisson channel \cite{lapidoth1998poisson}, \cite{aisha2016sum} and \cite{guo2008mutual}. The problem is set up such that there is a $4-$long binary input vector $X=[\allowbreak X_1, \allowbreak X_2,\allowbreak X_3, \allowbreak X_4]$ and each $X_i$ is a discrete random variable that assumes either of the two known values: $\lambda_0$ or $\lambda_1$ with probability $(1-p)$ and $p$, respectively. All $X_i$ are mutually independent and identically distributed. Conditioned on $X_i$, a Poisson process $\mathcal{P}_i(t)$ is initiated in continuous time $t$ \cite{ross1996stochastic}. It is known that If we count the arrivals for time $T_i $ from the conditional Poisson process we have another conditional counting Poisson process whose rate parameter at instant $T_i$ is $(T_i \cdot x_i)$. Hence we have, initially, four conditional Poisson processes: $\mathcal{P}_1(T_1 \cdot x_1)$; $\mathcal{P}_2(T_2 \cdot x_2)$; $\mathcal{P}_3(T_3 \cdot x_3)$ and $\mathcal{P}_4(T_4 \cdot x_4)$ depending on the realization that input vector $X$ assumes. Let ${4 \choose 2}$ be the set containing all possible pairs constituted from four elements of $X$. Summing elements of each of the $6-$pairs we then have another six conditional Poisson processes: $\mathcal{P}_5(T_5 \cdot (x_1+x_2))$; $\mathcal{P}_6(T_6 \cdot (x_1+x_3))$; $\mathcal{P}_7(T_7 \cdot (x_1+x_4))$; $\mathcal{P}_8(T_8 \cdot (x_2+x_3))$; $\mathcal{P}_9(T_9 \cdot (x_2+x_4))$ and $\mathcal{P}_{10}(T_{10} \cdot (x_3+x_4))$. Considering ${4 \choose 3}$ we have four processes:  $\mathcal{P}_{11}(T_{11} \cdot (x_1+x_2+x_3))$; $\mathcal{P}_{12}(T_{12} \cdot (x_1+x_2+x_4))$; $\mathcal{P}_{13}(T_{13} \cdot (x_1+x_3+x_4))$ and  $\mathcal{P}_{14}(T_{14} \cdot (x_2+x_3+x_4))$. Summing all the four components of $X$, we have $\mathcal{P}_{15}(T_{15} \cdot (x_1+x_2+x_3+x_4))$. Hence, there are $15-$ conditional point processes (in total) that we have to deal with to extract the maximum possible information or perform the best input signal detection by setting the counting times from $T_1$ to $T_{15}$ in a fixed given time $\sum_{i=1}^{15}=T$ as illustrated in fig. (\ref{f}).

The ideal way to address the problem would be to search for a solution in a $15- \allowbreak $ dimensional search-space by allowing $(T_1, T_2, \cdots T_{15})$ to have fifteen degrees-of-freedom. However, due to the computational complexity involved in exploring all fifteen dimensions we restrict ourselves to a reduced dimensional search-space, as said above. Therefore, we have considered only some special cases of time-configurations.
Four different types of time configurations are studied for each of the channels.  We call: \emph{individual} sensing when total given time $T$ is equally divided into $T_1=T_2=T_3=T_4=\frac{T}{4}$;  \emph{pair-wise} sensing when $T_5=T_6=T_7=T_8=T_9=T_{10}=\frac{T}{6}$;  \emph{triplets} sensing when $T_{11}=T_{12}=T_{13}=T_{14}=\frac{T}{4}$ and  \emph{joint} sensing when $T_{15}=T$.

The first problem is: does there exist a configuration among these four configurations which is always performing the best for any given time $T$ and prior $p$? To answer this we first fixed $p$, and then we consider $T$ as a free parameter and compute both the mutual information \cite{cover2012elements}, \cite{yeung2008information} and Bayes probability of total correct detections  \cite{schonhoff2006detection}, for a given set of parameters, and searched if there exist any instance for which one configuration is the best for some time and then another configuration becomes the best and so on. From mutual information perspective: it is computationally observed that when prior $p< 0.5$ then depending on the value of $T$ any of the four schemes can be better over the others however when $p \ge 0.5$ it is the individual sensing that works best. However, from the detection perspective this is not the case as indicated in fig. (\ref{f6}). It is further shown that in each configuration mutual information is concave in $T$.

The second problem: does there exist a \emph{hybrid} sensing mechanism that performs better than any of the above four configurations for fixed time $T$? A hybrid sensing is one when given time $T$ is divide into any one of the four sensing configurations and joint sensing according to the proportion:  $(1-\alpha) \cdot T $ and $\alpha \cdot T$ where $0 \le \alpha \le 1$, respectively. It turned out that if prior $p \ge 0.5$ then irrespective of other model parameters; individual sensing is the best among any other configurations. For $p$ close to zero hybrid sensing is better over any other as indicated in fig. (\ref{f1}). A concavity of mutual information w.r.t $\alpha$ is observed, but no proof is given.

For the vector Gaussian model we have a fixed unit covariance matrix and input $X$ only affects the mean vector. Replace all $T_i$ with $\sqrt{T_i}$ in Poisson model; we have the mean vector for Gaussian channel. It is found that triplet-sensing almost always outperforms any other configuration, irrespective of model parameters. This is shown in fig. (\ref{f7}) and fig. (\ref{f2}). It is shown that mutual information is concave in $T$ for any of the four configurations; further in hybrid sensing the mutual information remains concave in $\alpha$. However, Bayes probability of total correct detection is not necessarily consistent with mutual information results.

The paper is organized as follows: Section \ref{vectorchannel} defines the vector Poisson and Gaussian channel that we have considered. Section \ref{detd} describes the \emph{detection theoretic} model of the problem. Section \ref{comand} defines the computational setup. Finally, Section \ref{con} concludes the paper.

 \textit{Notation:} Upper case letters denote random vectors. Realizations of the random vectors are denoted by lower case letters. A number in subscript is used to show the component number of the random vector. We use $ X_1 $ and $ Y_1 $ to represent scalar input and output random variables, respectively. The superscript $ (\cdot)^{\intercal}  $ denotes the matrix/vector transpose. $ T $ is a given finite time. $ \alpha $ is an arbitrary positive scalar variable. $ \Phi $ represents the scaling matrix. $ p $ is the prior probability. $ f_X(x) $ denotes the probability mass function of $ X $. $\operatorname{Pois} (U;z)$ denotes the standard Poisson distribution of random variable $U$ with parameter $z$. We may omit $U$ in some cases. $K-$dimensional multivariate Gaussian distribution is represented by $\mathcal{N}(w;\mu,\Sigma )=(2 \pi)^{-\frac{k}{2}} \mathtt{det}(\Sigma)^{-\frac{1}{2}} e^{-{\frac{1}{2}}(w-\mu)^\intercal \Sigma^{-1}(w-\mu)}$. $w$ might be omitted for the purpose of  brevity.
\clearpage
\section{Vector Poisson and Gaussian Channels} \label{vectorchannel}
\subsection{Vector Poisson Channel} \label{vectorpoisson}
	We consider the vector Poisson channel model \cite{wang2014bregman}:
	\begin{IEEEeqnarray*}{rCl}
	\operatorname{Pois}(Y; \Phi X)&=&P_{Y|X}(Y|X)=\prod_{i=1}^{15} P_{Y_i|X}(Y_i|X)   \\ &=& \> \prod_{i=1}^{15} \operatorname{Pois}(Y_i;(\Phi X)_i)
		\label{E1} \IEEEyesnumber
	\end{IEEEeqnarray*} 	  
where
 $\operatorname{Pois} (U;z)$ denotes the standard Poisson distribution of random variable $U$ with parameter $z$.
 
 We assume input $X=(X_1,X_2,X_3,X_4) \in \{ \lambda_0, \lambda_1  \}^{4}$ such that $0 \le \lambda_0< \lambda_1$, each $X_i$ is independent and identically distributed with a pmf: $p_{X_i}(x_i=\lambda_0)=1-p$ and $p_{X_i}(x_i=\lambda_1)=p$. $Y=(Y_1,Y_2, \cdots Y_{15}) \in \mathbb{Z}_{+}^{15}$ and
 	\begin{equation}
 	\label{E3} 
 	\begin{aligned}
 \Phi
 		&=
 		\begin{bmatrix}
 			T_1 & 0  & 0  & 0   \\
 			0 & T_2  & 0  & 0 \\
 			0 & 0 &  T_3  & 0 \\
 			0  & 0 & 0  & T_4 \\
 			T_5 & T_5  & 0  & 0 \\
 			T_6  & 0 & T_6  & 0 \\
 			T_7  & 0  & 0 & T_7 \\
 			0  &  T_8 & T_8  & 0 \\
 			0  &  T_9 & 0   & T_9 \\
 			0  &  0   & T_{10} & T_{10} \\
 			T_{11} & T_{11} & T_{11} &0\\
 			T_{12} & T_{12} &0 & T_{12} \\
 			T_{13} & 0 & T_{13} & T_{13} \\
 			0 & T_{14} & T_{14} & T_{14} \\
 			T_{15} & T_{15} & T_{15} & T_{15}\\
 		\end{bmatrix}.
 	 	 	\end{aligned}
 \end{equation}

The conditional distribution of vector $Y$ given $X$ is a multivariate Poisson distribution:  
\begin{equation}
	\label{E2} 
	\begin{aligned}
		Y\Bigg|\Big(X=(x_1 \: x_2 \: x_3 \: x_4)\Big)
		& \sim
		\operatorname{Pois} \Bigg(\begin{bmatrix}
			T_1 \cdot x_1   \\
			T_2 \cdot x_2 \\
			T_3 \cdot x_3 \\
	    	T_4 \cdot x_4 \\
			T_5 \cdot (x_1+x_2) \\
			T_6 \cdot (x_1+x_3) \\
			T_7 \cdot (x_1+x_4) \\
			T_8 \cdot (x_2+x_3) \\
			T_9 \cdot (x_2+x_4) \\
			T_{10} \cdot (x_3+x_4) \\
			T_{11} \cdot (x_1+x_2+x_3) \\
			T_{12} \cdot (x_1+x_2+x_4) \\
			T_{13} \cdot (x_1+x_3+x_4) \\
			T_{14} \cdot (x_2+x_3+x_4) \\
			T_{15} \cdot (x_1+x_2+x_3+x_4) \\
		\end{bmatrix}	
		\Bigg),\\
	\end{aligned}
\end{equation}
We define mutual information $I(X;Y)$ as
\begin{IEEEeqnarray*}{rCl}
	I(X;Y)=H(Y)- H(Y|X), \label{AL00} \IEEEyesnumber
\end{IEEEeqnarray*}
where $H(Y)$ is an entropy of a finite Poisson mixture model given as
\begin{IEEEeqnarray*}{rCl}
	H(Y)	&=&  - \sum_{y_1=-\infty}^{\infty}\sum_{y_2=-\infty}^{\infty} \cdots \cdots \sum_{y_{15}=-\infty}^{\infty}\: P(Y) \cdot \operatorname{Log}_2[P(Y)] \label{AL70} \IEEEyesnumber
\end{IEEEeqnarray*}
where
\begin{IEEEeqnarray*}{rCl}
	P(Y)	&=& \sum_{i=1}^{16} \Bigg( P_{X_i}(X_i=x_1,x_2,x_3,x_4) \cdot  \operatorname{Pois}  \Big(y_1;{T_1} x_1 \Big) \cdot  \\ && \> \operatorname{Pois}  \Big(y_2;{T_2} x_2 \Big) \cdot  \operatorname{Pois}  \Big(y_3;{T_3} x_3 \Big)  \cdot \operatorname{Pois}  \Big(y_4;{T_4}x_4 \Big) \cdot \operatorname{Pois}  \Big(y_5;{T_5} (x_1+x_2) \Big)  \cdot
	\\ && \>\operatorname{Pois}  \Big(y_6;{T_6}  (x_1+x_3) \Big) \cdot
	\operatorname{Pois}  \Big(y_7;{T_7}  (x_1+x_4) \Big) \cdot
	\operatorname{Pois}  \Big(y_8;{T_8} (x_2+x_3) \Big) \cdot
	\\ && \> \operatorname{Pois}  \Big(y_9;{T_9}(x_2+x_4) \Big) \cdot
	\operatorname{Pois}  \Big(y_{10};{T_{10}} (x_3+x_4) \Big) \cdot
	\operatorname{Pois}  \Big(y_{11};{T_{11}} (x_1+x_2+x_3) \Big) \cdot
	\\ && \> \operatorname{Pois}  \Big(y_{12};{T_{12}}(x_1+x_2+x_4) \Big) \cdot
	\operatorname{Pois}  \Big(y_{13};{T_{13}}(x_1+x_3+x_4) \Big) \cdot
	\\ && \> \operatorname{Pois}  \Big(y_{14};{T_{14}} (x_2+x_3+x_4) \Big) \cdot
	\operatorname{Pois}  \Big(y_{15};{T_{15}} (x_1+x_2+x_3+x_4) \Big)	\Bigg), \label{AL80} \IEEEyesnumber
\end{IEEEeqnarray*}
and
\begin{IEEEeqnarray*}{rCl}
	H(Y|X)	&=&-  \sum_{y_1=-\infty}^{\infty}\sum_{y_2=-\infty}^{\infty} \cdots \cdots \sum_{y_{15}=-\infty}^{\infty} \sum_{i=1}^{16} P_{X_i}(X_i) \cdot P(Y|X_i) \cdot \operatorname{Log}_2[P(Y|X_i)] \label{AL90} \IEEEyesnumber
\end{IEEEeqnarray*}
where $P(Y|X_i) =\prod_{j=1}^{15}\operatorname{Pois}(Y_j;(\Phi X)_i)$.
\begin{theorem}
	$I(X_1,X_2,X_3,X_4;Y_1,Y_2,Y_3 \cdots Y_{15})$ is symmetric in variable-groups: $\allowbreak (T_1, \allowbreak T_2, \allowbreak T_3,\allowbreak T_4)\allowbreak $; $\allowbreak  (T_5, T_6, T_7,\allowbreak  T_8, T_9,T_{10}) $; and  $ (T_{11}, T_{12}, T_{13},T_{14}) $.
\end{theorem}
\begin{IEEEproof}
	Mutual information $I(X;Y)$ given in (\ref{AL00}) is invariant under any permutation of variables belonging to the same group. That means interchanging the variables within the same group leaves the expression unchanged.
\end{IEEEproof}
\subsubsection{Unconstrained objective} \label{u2}
For a vector Poisson channel with given prior $p$, $\lambda_0$ and $\lambda_1$, which of the following four methods are better over the others when each expression is a function of $T$ solely, 
\begin{IEEEeqnarray}{lCr}
	I(X_1, X_2, X_3, X_4 ; \overbrace{Y_1, Y_2, Y_3, Y_4)}^{\rm{4-Singlets}}    & \text{s.t.}  & T_1=T_2=T_3=T_4=\frac{T}{4}	\IEEEnonumber
	\IEEEnonumber \\ \label{e21} \IEEEyesnumber \\
	I(X_1, X_2, X_3, X_4 ;  \overbrace{Y_5, Y_6, Y_7, Y_8, Y_9, Y_{10})}^{\rm{6-Pairs}}  \quad & \text{s.t.} & \quad T_5=T_6=T_7=T_8=T_9=T_{10}=\frac{T}{6}	
	\IEEEnonumber \\ \label{e22} \IEEEyesnumber \\
	I(X_1, X_2, X_3, X_4 ;  \overbrace{Y_{11}, Y_{12},Y_{13}, Y_{14})}^{\rm{4-Triplets}}  \quad & \text{s.t.} & \quad T_{11}=T_{12}=T_{13}=T_{14}=\frac{T}{4}	
	\IEEEnonumber \\ \label{e23}  \IEEEyesnumber \\
	I(X_1, X_2, X_3, X_4 ;  \overbrace{Y_{15})}^{\rm{1-Quadruplet}}  \quad & \text{s.t.} & \quad T_{15}=T
	\IEEEnonumber \\ \label{e24}  \IEEEyesnumber
\end{IEEEeqnarray}
\begin{theorem}
	$I(X_1,X_2,X_3,X_4;Y_5,Y_6,Y_7,Y_8,Y_9,Y_{10})$ $\text{s.t. }$  $T_5=T_6=T_7=T_8=T_9=T_{10}=\frac{T}{6}$	   is concave in $T$.
\end{theorem}
\begin{IEEEproof}
	$I(X_1,X_2,X_3,X_4;Y_5,Y_6,Y_7,Y_8,Y_9,Y_{10})= \allowbreak  I((X_1+X_2),(X_1+X_3),\allowbreak (X_1+X_4),\allowbreak (X_2+\allowbreak X_3),\allowbreak(X_2+\allowbreak X_4),\allowbreak(X_3+X_4);Y_5,Y_6,Y_7,Y_8,Y_9,Y_{10})$. 
	
	Consider the R.H.S of the above equation. From \cite[p.~1315]{atar2012mutual}, for a random n-tuple vector $X=(X_1, \cdots X_n)$ and for $T \ge 0$, let $Y=(Y_1, \cdots Y_n)$ be jointly distributed with $X$ such that given $X$, the components of $Y$ are independent with $Y_i|X \sim \operatorname{Poiss}(T \cdot X_i), 1 \le i \le n.$ Then $\operatorname{mmle}(T)=E \Big[ \sum_{i=1}^{n} l(X_i,E[X_i | Y])   \Big].$ Since $\frac{\partial}{\partial T} I(X_i;Y) = \operatorname{mmle}(T).$ Each $E \Big[  l(X_i,E[X_i | Y])\Big]  $ is concave in $T$. Sum of concave functions result in another Concave function.
\end{IEEEproof}
\begin{corollary}
	Expressions in (\ref{e21}),	(\ref{e23}) and (\ref{e24}) are concave in $T$ too. 
\end{corollary}
Note that expressions in (\ref{e21}), (\ref{e22}) and (\ref{e23}), have a tight upper bound of $H(X_1,X_2,X_3,X_4)$ as $T\rightarrow \infty $ since the corresponding mappings: from $X$ to $6-$pairs and from $X$ to $4-$triplets are invertible. Whereas, the expression (\ref{e24}) has a tight upper bound of  $H(\sum_{i=1}^{4} X_i)$ when $T \rightarrow \infty,$ the mapping from $(X_1,X_2,X_3,X_4) \mapsto \sum_{i=1}^{4} X_i$ is non-invertible.
\subsubsection{Constraint objective} \label{p2}
The second objective is to determine which of the following three configurations are better over the others for a given prior $p$,  $\lambda_0$, $\lambda_1$ and given fixed time $T$ i.e.,
\begin{IEEEeqnarray*}{lCl}
	\mathrm{Config-1} : &&  I(X; \overbrace{Y_1, Y_2, Y_3, Y_4,Y_{15})}^{\rm{4-Singlets+1-Quadruplet}} \: \text{s.t.} \: T_1=T_2=T_3=T_4=\frac{T-\alpha}{4}, T_{15}=\alpha. 		\label{e26}  \IEEEyesnumber \\
	\mathrm{Config-2} : && I(X ; \overbrace{Y_5, Y_6, Y_7, Y_8, Y_9, Y_{10},Y_{15})}^{\rm{6-Pairs+1-Quadruplet}} \: \text{s.t.} \: T_5=T_6=T_7=T_8=T_9=T_{10}=\frac{T-\alpha}{6}, \\ && \> T_{15}=\alpha. 	\label{e27} \IEEEyesnumber \\
	\mathrm{Config-3:} && I(X ; \overbrace{Y_{11}, Y_{12},Y_{13}, Y_{14},Y_{15})}^{\rm{4-Triplets+1-Quadruplet}}   \: \text{s.t.} \: T_{11}=T_{12}=T_{13}=T_{14}=\frac{T-\alpha}{4}, T_{15}=\alpha. \IEEEnonumber \\
	\label{e28}  \IEEEyesnumber
\end{IEEEeqnarray*}
where $0 \le \alpha \le T$.

\subsection{Vector Gaussian Channel} \label{vectorgaussian}
  We consider the vector Gaussian channel model as defined in \cite{wang2014bregman}  i.e., $Y|X \sim \mathcal{N}(\Phi X, \mathrm{I})$. For a scalar Gaussian channel $Y=\sqrt{T} \cdot X+N$ with $N \sim \mathcal{N}(0,1)$; $I(X;Y)$ is concave in $T$ for arbitrary input signalling \cite{guo2005mutual}. We extend this scalar model to the vector case. We may also write as	
	\begin{equation}
		\label{e0} 
		\begin{aligned}
			\begin{bmatrix}
				Y_1     \\
				Y_2  \\
				Y_3   \\
					Y_4     \\
				Y_5  \\
				Y_6   \\
					Y_7    \\
				Y_8  \\
				Y_9   \\
					Y_{10}     \\
				Y_{11} \\
				Y_{12} \\
					Y_{13}    \\
				Y_{14} \\
				Y_{15}  \\
			\end{bmatrix}
			&=
			\begin{bmatrix}
				\sqrt{T_1} & 0  & 0  & 0   \\
				0 & \sqrt{T_2}  & 0  & 0 \\
				0 & 0 &  \sqrt{T_3}  & 0 \\
				0  & 0 & 0  & \sqrt{T_4} \\
				\sqrt{T_5} & \sqrt{T_5}  & 0  & 0 \\
				\sqrt{T_6}  & 0 & \sqrt{T_6}  & 0 \\
				\sqrt{T_7}  & 0  & 0 & \sqrt{T_7} \\
				 0  &  \sqrt{T_8} & \sqrt{T_8}  & 0 \\
				 0  &  \sqrt{T_9} & 0   & \sqrt{T_9} \\
				 0  &  0   & \sqrt{T_{10}} & \sqrt{T_{10}} \\
				 \sqrt{T_{11}} & \sqrt{T_{11}} & \sqrt{T_{11}} &0\\
				 \sqrt{T_{12}} & \sqrt{T_{12}} &0 & \sqrt{T_{12}} \\
				 \sqrt{T_{13}} & 0 & \sqrt{T_{13}} & \sqrt{T_{13}} \\
				 0 & \sqrt{T_{14}} & \sqrt{T_{14}} & \sqrt{T_{14}} \\
				 \sqrt{T_{15}} & \sqrt{T_{15}} & \sqrt{T_{15}} & \sqrt{T_{15}}\\
			\end{bmatrix} 
			\begin{bmatrix}
				X_1   \\
				X_2   \\
				X_3   \\
				X_4   \\
			\end{bmatrix} +
			\begin{bmatrix}
				N_1  \\
				N_2  \\
				N_3 \\
				N_4  \\
				N_5  \\
				N_6\\
				N_7  \\
				N_8  \\
				N_9 \\
				N_{10}  \\
				N_{11}  \\
				N_{12} \\
				N_{13}\\
				N_{14}\\
				N_{15} \\
			\end{bmatrix},\\
		\end{aligned}
	\end{equation}	
where each $X_i$ is independent and identical distributed (i.i.d) discrete random variable with support $\{\lambda_0, \lambda_1 \}$ such that $ p $ is the probability of occurrence of $ \lambda_1 $ and $(1-p)$ is the occurrence of  $ \lambda_0 $ i.e., probability mass function $ f $ of scalar input random variable $ X_i$  is
\begin{IEEEeqnarray*}{rCl}
	f_{X_i}(x_i)=\left\{
	\begin{array}{ll}
		p & \quad x  =  \: \lambda_1 \\ 
		(1-p) & \quad x  =  \: \lambda_0  \\    
	\end{array}
	\right. \label{w8} \IEEEyesnumber
\end{IEEEeqnarray*}
Noise vector $N$ is a multivariate Gaussian with zero mean and identity covariance matrix $\mathrm{I}$; and independent of input $X$. The constraint on the scaling matrix is $\sum_{i=1}^{15} T_i=T$. 
The conditional distribution of vector $Y$ given $X$ is a multivariate Gaussian:
\begin{equation}
	\label{e01} 
	\begin{aligned}
		Y\Bigg|\Big(X=(x_1 \: x_2 \: x_3 \: x_4)\Big)
		& \sim
		\mathcal{N} \Bigg(\begin{bmatrix}
			\sqrt{T_1} \cdot x_1   \\
			\sqrt{T_2} \cdot x_2 \\
			\sqrt{T_3} \cdot x_3 \\
			\sqrt{T_4} \cdot x_4 \\
			\sqrt{T_5} \cdot (x_1+x_2) \\
			\sqrt{T_6} \cdot (x_1+x_3) \\
			\sqrt{T_7} \cdot (x_1+x_4) \\
			\sqrt{T_8} \cdot (x_2+x_3) \\
			\sqrt{T_9} \cdot (x_2+x_4) \\
			\sqrt{T_{10}} \cdot (x_3+x_4) \\
			\sqrt{T_{11}} \cdot (x_1+x_2+x_3) \\
			\sqrt{T_{12}} \cdot (x_1+x_2+x_4) \\
			\sqrt{T_{13}} \cdot (x_1+x_3+x_4) \\
			\sqrt{T_{14}} \cdot (x_2+x_3+x_4) \\
			\sqrt{T_{15}} \cdot (x_1+x_2+x_3+x_4) \\
		\end{bmatrix},
\mathrm{I}
\Bigg),\\
	\end{aligned}
\end{equation}
where $\mathrm{I}$ is an identity matrix of size $15 \times 15 $.



We define mutual information $I(X;Y)$ as
\begin{IEEEeqnarray*}{rCl}
	I(X;Y)=H(Y)- H(Y|X), \label{AL0} \IEEEyesnumber
\end{IEEEeqnarray*}
where $H(Y)$ is a differential entropy of a finite Gaussian mixture model (gmm) given as
\begin{IEEEeqnarray*}{rCl}
	H(Y)	&=&  - \int_{y_1=-\infty}^{\infty}\int_{y_2=-\infty}^{\infty} \cdots \cdots \int_{{y_{15}}=-\infty}^{\infty}\: P(Y) \cdot \operatorname{Log}_2[P(Y)] \: \,d{y_1}\,d{y_2}\,d{y_3} \cdots \cdots \,d{y_{15}}, \label{AL7} \IEEEyesnumber
\end{IEEEeqnarray*}
where
\begin{IEEEeqnarray*}{rCl}
	P(Y)	&=& \sum_{i=1}^{16} \Bigg( P_{X_i}(X_i=x_1,x_2,x_3,x_4) \cdot  \mathcal{N} \Big(y_1;\sqrt{T_1} x_1, 1 \Big) \cdot  \\ && \> \mathcal{N} \Big(y_2;\sqrt{T_2} x_2, 1 \Big) \cdot  \mathcal{N} \Big(y_3;\sqrt{T_3} x_3, 1 \Big)  \cdot \mathcal{N} \Big(y_4;\sqrt{T_4}x_4, 1 \Big) \cdot \mathcal{N} \Big(y_5;\sqrt{T_5} (x_1+x_2), 1 \Big)  \cdot
 	\\ && \>\mathcal{N} \Big(y_6;\sqrt{T_6}  (x_1+x_3), 1 \Big) \cdot
 \mathcal{N} \Big(y_7;\sqrt{T_7}  (x_1+x_4), 1 \Big) \cdot
 \mathcal{N} \Big(y_8;\sqrt{T_8} (x_2+x_3), 1 \Big) \cdot
	\\ && \> \mathcal{N} \Big(y_9;\sqrt{T_9}(x_2+x_4), 1 \Big) \cdot
 \mathcal{N} \Big(y_{10};\sqrt{T_{10}} (x_3+x_4), 1 \Big) \cdot
 \mathcal{N} \Big(y_{11};\sqrt{T_{11}} (x_1+x_2+x_3), 1 \Big) \cdot
	\\ && \> \mathcal{N} \Big(y_{12};\sqrt{T_{12}}(x_1+x_2+x_4), 1 \Big) \cdot
 \mathcal{N} \Big(y_{13};\sqrt{T_{13}}(x_1+x_3+x_4), 1 \Big) \cdot
	\\ && \> \mathcal{N} \Big(y_{14};\sqrt{T_{14}} (x_2+x_3+x_4), 1 \Big) \cdot
 \mathcal{N} \Big(y_{15};\sqrt{T_{15}} (x_1+x_2+x_3+x_4), 1 \Big)	\Bigg), \label{AL8} \IEEEyesnumber
\end{IEEEeqnarray*}
and
\begin{IEEEeqnarray*}{rCl}
	H(Y|X)	&=& 15 \times 0.5 \times \operatorname{Log}_2[2 \pi \: e ]. \label{AL9} \IEEEyesnumber
\end{IEEEeqnarray*}
The multidimensional integral defined in (\ref{AL7}) have no closed-form solution, and therefore we need to resort to the Monte Carlo method. The following method is used to numerically evaluated the integral using sampling from a finite Gaussian mixture.
\begin{IEEEeqnarray*}{rCl}
H(Y)&=&E[-\operatorname{Log}_2[P_Y(Y)]]  \\ &\approx& \>-\frac{\sum_i \operatorname{Log}_2[P_Y(s_i)]}{N_s},
\label{AL4} \IEEEyesnumber
\end{IEEEeqnarray*} 	  
Where $P_Y(\cdot)$ is the mixture probability distribution of $Y$, $N_s$ is the number of MC samples and $s_i$ is the $i^{th}$ sample from multivariate Gaussian mixture distribution.
\begin{theorem}
	$I(X_1,X_2,X_3,X_4;Y_1,Y_2,Y_3 \cdots Y_{15})$ is symmetric in variable-groups: $\allowbreak (T_1, \allowbreak T_2, \allowbreak T_3,\allowbreak T_4)$; $ \allowbreak (T_5, T_6, T_7,\allowbreak  T_8, T_9,T_{10}) $; and  $ (T_{11}, T_{12}, T_{13},T_{14}) $.
\end{theorem}
\begin{IEEEproof}
	Mutual information $I(X;Y)$ given in (\ref{AL0}) is invariant under any permutation of variables belonging to the same group. That means interchanging the variables within the same group leaves the expression unchanged.
\end{IEEEproof}
\subsubsection{Unconstrained objective} \label{u3}
For a vector Gaussian channel with given prior $p$, $\lambda_0$ and $\lambda_1$, which of the following four methods are better over the others when each expression is a function of $T$ solely, 
\begin{IEEEeqnarray*}{lCr}
	I(X_1, X_2, X_3, X_4 ; \overbrace{Y_1, Y_2, Y_3, Y_4}^{\rm{4-Singlets}})   & \text{s.t.}  & T_1=T_2=T_3=T_4=\frac{T}{4}	\\
	\label{e41} \IEEEyesnumber \\
	I(X_1, X_2, X_3, X_4 ; \overbrace{Y_5, Y_6, Y_7, Y_8, Y_9, Y_{10}}^{\rm{6-Pairs}}) \quad & \text{s.t.} & \quad T_5=T_6=T_7=T_8=T_9=T_{10}=\frac{T}{6}	\\
	\label{e42} \IEEEyesnumber \\
	I(X_1, X_2, X_3, X_4 ; \overbrace{Y_{11}, Y_{12},Y_{13}, Y_{14}}^{\rm{4-Triplets}}) \quad & \text{s.t.} & \quad T_{11}=T_{12}=T_{13}=T_{14}=\frac{T}{4}	\\
	\label{e43} \IEEEyesnumber \\
	I(X_1, X_2, X_3, X_4 ; \overbrace{Y_{15}}^{\rm{1-Quadruplet}}) \quad & \text{s.t.} & \quad T_{15}=T	\\
	\label{e44} \IEEEyesnumber
\end{IEEEeqnarray*}
\begin{theorem}
	$ I(X_1,X_2,X_3,X_4 ; Y_5, Y_6, Y_7, Y_8, Y_9, Y_{10}) $ in (\ref{e42}) is concave in $T$. 
\end{theorem}
\begin{IEEEproof}
	It is noted in \cite[Theorem 5]{payaro2009hessian} that mutual information is a concave function of the squared singular values $(\boldsymbol{\lambda})$ of the precoder matrix $P$ if the first $m^{'}$ eigenvectors of the channel covariance matrix $(\mathbf{R}_{H}=H^\intercal \mathbf{R}_{Z}^{-1} H)$ coincide with the left singular vectors of the precoder $P$ i-e $\mathsf{H}_\lambda I(S;Y) \le 0$ for the signal model $Y=HPS+Z$ where $H \in \mathbb{R}^{n \times p}$ is the channel, $S$ is the input signaling $S \in \mathbb{R}^{m}$, $P$ is a precoder matrix $P \in \mathbb{R}^{p \times m}$ and $Z \in  \mathbb{R}^{n}$ is Gaussian noise independent of the input $S$ and has covariance matrix $\mathbf{R}_Z$.
	
	For our problem: $H= I,$ $\mathbf{R}^{-1}_Z=\Lambda=I,$ $P=\Phi$, $S=X$ and $Z=N.$ The singular value decomposition of $\Phi=U \Sigma V^*$. 
	
	By substituting $T_1=T_2=T_3=T_4=0$, $T_5=T_6=\cdots T_{10}=T$ and $T_{15}=0$ in (\ref{e0}), the squared singular values of $\Phi$ are $[ \lambda_1, \lambda_2 \cdots \lambda_{15}]\allowbreak = \allowbreak [6T, 2T, 2T, 2T, 0, 0, 0, \cdots 0]$.  This is just the composition with an affine transformation on the domain.
\end{IEEEproof}
\begin{corollary}
	$  I(X; Y_1, Y_2, Y_3, Y_4) $, $I(X ; Y_{11}, Y_{12},Y_{13}, Y_{14})$ and  $  I(X; Y_{15}) $ are concave in $T$, since squared singular values are $ \allowbreak [ T, T, T, T, 0, 0 \cdots 0]$, $ \allowbreak  [9T, 2T, \allowbreak2T,\allowbreak 2T, \allowbreak0, 0, 0, \allowbreak \cdots 0]$ and $ \allowbreak  [4T, \allowbreak0, 0, 0, \allowbreak \cdots 0]$ respectively.
\end{corollary}
\subsubsection{Constraint objective} \label{c2}
The second objective for the vector Gaussian channel is which of the following three configurations are better over the others for a given prior $p$,  $\lambda_0$, $\lambda_1$ and given fixed time $T$ i.e.,
\small
\begin{IEEEeqnarray*}{lCl}
	\mathrm{Config-1} : &&  I(X; \overbrace{Y_1, Y_2, Y_3, Y_4,Y_{15})}^{\rm{4-Singlets+1-Quadruplet}} \: \text{s.t.} \: T_1=T_2=T_3=T_4=\frac{T-\alpha}{4}, T_{15}=\alpha. 		
	\label{e66} \IEEEyesnumber \\
	\mathrm{Config-2} : && I(X ; \overbrace{Y_5, Y_6, Y_7, Y_8, Y_9, Y_{10},Y_{15})}^{\rm{6-Pairs+1-Quadruplet}} \: \text{s.t.} \: T_5=T_6=T_7=T_8=T_9=T_{10}=\frac{T-\alpha}{6}, \\ && \> T_{15}=\alpha. 	\label{e67} \IEEEyesnumber \\
	\mathrm{Config-3:} && I(X ; \overbrace{Y_{11}, Y_{12},Y_{13}, Y_{14},Y_{15})}^{\rm{4-Triplets+1-Quadruplet}}   \: \text{s.t.} \: T_{11}=T_{12}=T_{13}=T_{14}=\frac{T-\alpha}{4}, T_{15}=\alpha. 
	\label{e68} \IEEEyesnumber
\end{IEEEeqnarray*}
where $0 \le \alpha \le T$.
\normalsize
\begin{theorem}
$ I(X_1,X_2,X_3,X_4 ; Y_5, Y_6, Y_7, Y_8, Y_9, Y_{10},Y_{15}) \: \text{ s.t. } \: T_5=\allowbreak T_6\allowbreak =T_7\allowbreak =T_8= \allowbreak T_9 \allowbreak =T_{10}= \allowbreak \frac{T-\alpha}{6}, \allowbreak  T_{15}=\alpha, $ where $ 0 \le \alpha \le T, $ is concave in $\alpha$. 
\end{theorem}
\begin{IEEEproof}
We again resort to the \cite[Theorem 5]{payaro2009hessian}.

For our problem: $H= I,$ $\mathbf{R}^{-1}_Z=\Lambda=I,$ $P=\Phi$, $S=X$ and $Z=N.$ The singular value decomposition of $\Phi=U \Sigma V^*$. 

By substituting $T_1=T_2=T_3=T_4=0$, $T_5=T_6=\cdots T_{10}=\frac{T-\alpha}{2}$ and $T_{15}=\alpha$ in (\ref{e0}), the squared singular values of $\Phi$ are $[ \lambda_1, \lambda_2 \cdots \lambda_{15}]\allowbreak = \allowbreak [ 0, 0 \cdots 0, \frac{T-\alpha}{3},\allowbreak \frac{T-\alpha}{3},\allowbreak \frac{T-\alpha}{3},\allowbreak T+3 \alpha]$.  This is just the composition with an affine transformation on the domain. Concavity remains preserved under affine transformation \cite[page 79-86]{boyd2004convex}.
\end{IEEEproof}
\begin{corollary}
$  I(X; Y_1, Y_2, Y_3, Y_4,Y_{15}) \: \text{ s.t. } \: T_1= \allowbreak T_2= \allowbreak T_3= \allowbreak T_4= \allowbreak \frac{T-\alpha}{4}, T_{15}=\alpha$ and $I(X ; Y_{11},\allowbreak Y_{12},Y_{13},\allowbreak Y_{14},Y_{15})   \: \text{ s.t. } \: T_{11}=T_{12}=T_{13}=T_{14}=\frac{T-\alpha}{4}, T_{15}=\alpha$ are concave in $\alpha$, since squared singular values are $[ \lambda_1, \lambda_2 \cdots \lambda_{15}]\allowbreak = \allowbreak [ 0, 0 \cdots 0, \frac{T-\alpha}{4},\frac{T-\alpha}{4},\frac{T-\alpha}{4},\frac{T+15 \alpha}{4}]$ and $[ \lambda_1, \lambda_2 \cdots \lambda_{15}]\allowbreak = \allowbreak [ 0, 0 \cdots 0, \frac{T-\alpha}{4},\allowbreak \frac{T-\alpha}{4},\allowbreak \frac{T-\alpha}{4},\allowbreak \frac{9T+7 \alpha}{4}]$, respectively.
\end{corollary}
Therefore, the constraint objectives in (\ref{e66}), (\ref{e67}) and (\ref{e68}) are all concave in $\alpha$.
\section{Detection Theoretic Description} \label{detd}

\subsection{Bayes criterion} \label{adoptcri}
In terms of Bayes detection we may consider the problem as deciding among the $16-$ hypotheses $(\mathcal{H}_i, 1 \le i \le 16)$ for a fixed time-proportions. Considering the prior probability of each hypothesis as $\pi_i$ such that $\sum_{i=1}^{16} \pi_i=1$. Let $C_{il}$ is the cost of deciding $\mathcal{D}_i$ when $\mathcal{H}_l$ is correct, then the average cost is $r=\sum_{i=1}^{16} \sum_{l=1}^{16} \pi_l C_{il} P_{il}$, where $P_{il}$ is the probability of deciding $\mathcal{D}_i$ when $\mathcal{H}_l$ is true.

For Gaussian problem with fixed sensing-time proportions; 
$\mathcal{H}_i : \mathcal{N}_{15}(\mu_i,\mathbb{I}),$ with prior $\pi_i, \quad 1 \le i \le 16.$
 Where $\mathcal{N}_{15}(\mu_i,\mathbb{I})$ is a $15-$dimensional multivariate normal distribution with fixed covariance unit-matrix and $15-$component random mean vector $\mu_i$.
 $P_{il}= \int_{y \in \mathcal{R}_i}^{} P_l (y_1,y_2 \cdots y_{15}|\mathcal{H}_i) \partial y.$
We only consider the MAP criterion where cost is
\begin{IEEEeqnarray*}{rCl}
C_{il}=\left\{
	\begin{array}{ll}
		0 & \quad i  =  l \\ 
     	1 & \quad i  \ne l \\    
	\end{array}
	\right. \label{e10} \IEEEyesnumber
\end{IEEEeqnarray*}
This simplifies the detection rule to deciding:
\begin{IEEEeqnarray*}{rCl}
\mathcal{D}_i: \quad  \text{if} \quad
\pi_i p_i(y_1,y_2, \cdots y_{15} | \mathcal{H}_i) \allowbreak \ge \pi_n p_n(y_1,y_2, \cdots y_{15} | \mathcal{H}_n) \quad \text{for all} \quad n \ne i;
\label{e11} \IEEEyesnumber
\end{IEEEeqnarray*}
 and for any fixed time-proportions $(T_1,T_2, \cdots T_{15})$ under consideration.
 
For Poisson problem with fixed sensing-times:
$\mathcal{H}_i : \operatorname{Poiss}_{15}(\mu_i),$ with prior $\pi_i, \quad 1 \le i \le 16.$
Where $\operatorname{Pois}_{15}(\mu_i)=  \prod_{i=1}^{15} \operatorname{Pois}((\Phi X)_i) $ is a $15-$dimensional multivariate Poisson distribution with $15-$ component random mean vector $\mu_i$.
$P_{il}= \sum_{y \in \mathcal{R}_i}^{} P_l (y_1,y_2 \cdots y_{15}|\mathcal{H}_i)$.
Thus we are interested in minimizing the Bayes risk $ r $(under both constrained and unconstrained objectives defined above) for any given structure in  time-proportions $ (T_1,T_2 \cdots T_{15})$ i-e
\begin{IEEEeqnarray*}{rCl}
	& \underset{(T_1,T_2, \cdots T_{15})}{\text{min}} \: r \quad \text{s.t.} \: \:   \sum_{i=1}^{15} T_i=T. \label{e4} \IEEEyesnumber
\end{IEEEeqnarray*} 
Equivalently, we may write
\begin{IEEEeqnarray*}{rCl}
	& \underset{(T_1,T_2, \cdots T_{15})}{\text{max}} \: P_d \quad  \text{s.t.} \: \:  \sum_{i=1}^{15} T_i=T \label{e8} \IEEEyesnumber
\end{IEEEeqnarray*}
where $ P_d $ is probability of total correct detections, $ P_d=1-r. $

\begin{conjecture}
The optimal solution of finding the best time-proportions in a given fixed time $T$, both under information theoretic and detection theoretic metrics, has a specific structure:  $(T_1:=a,T_2:=a,T_3:=a,T_4:=a,T_5:=b,T_6:=b,T_7:=b,T_8:=b,T_9:=b,T_{10}:=b,T_{11}:=c,T_{12}:=c,T_{13}:=c,T_{14}:=c,T_{15}:=d)$. Where $0 \le a,b,c,d \le T$ and $ 4 \cdot a + 6 \cdot b + 4 \cdot c + d =T$.
\end{conjecture}
\section{Computational setup} \label{comand}
\begin{figure}[ht]
	\begin{subfigure}{.49\textwidth}
		\includegraphics[width=\linewidth]{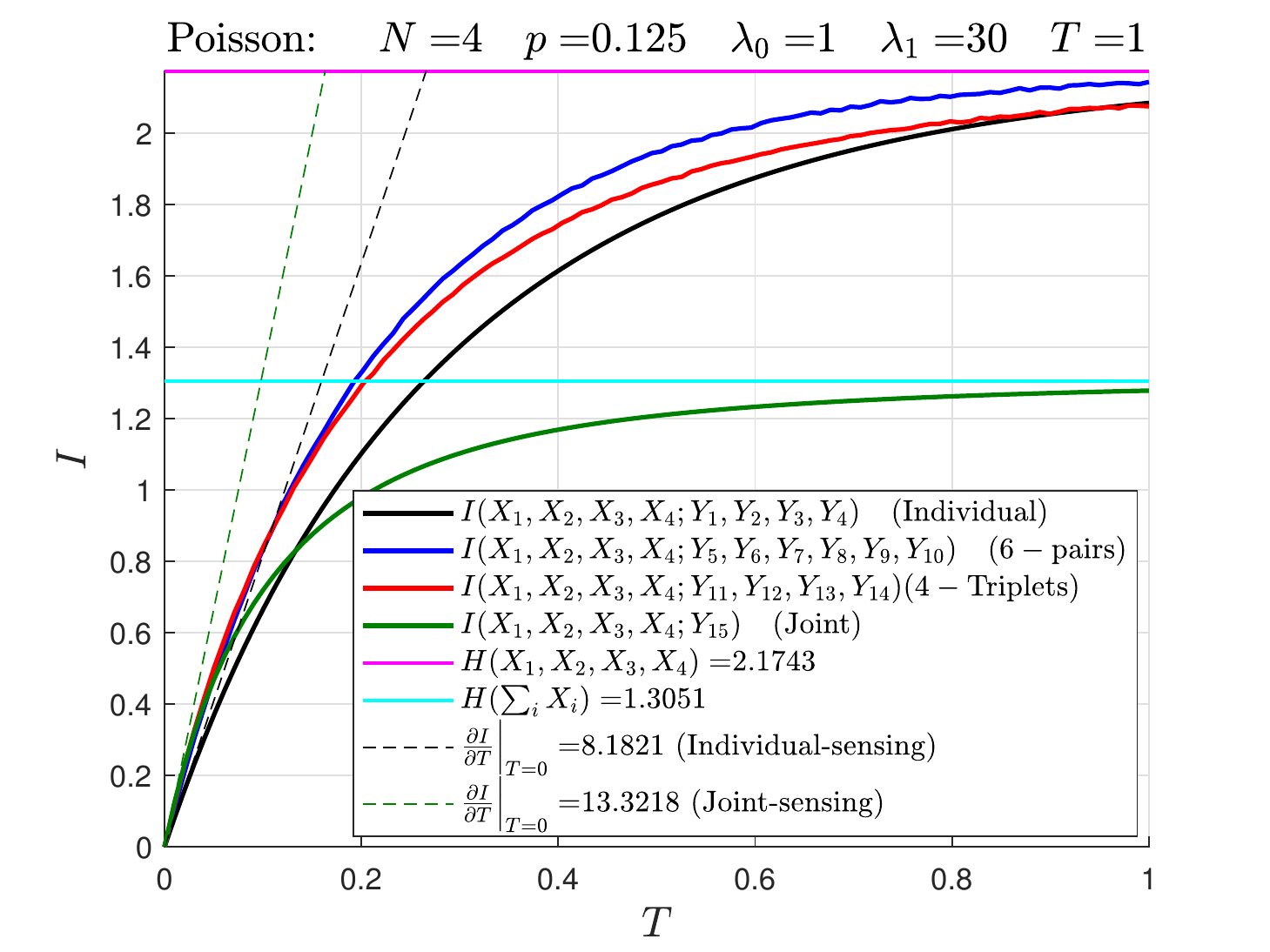}
		\caption{ }
		\label{fig6a}
	\end{subfigure}
	\begin{subfigure}{.49\textwidth}
		\includegraphics[width=\linewidth]{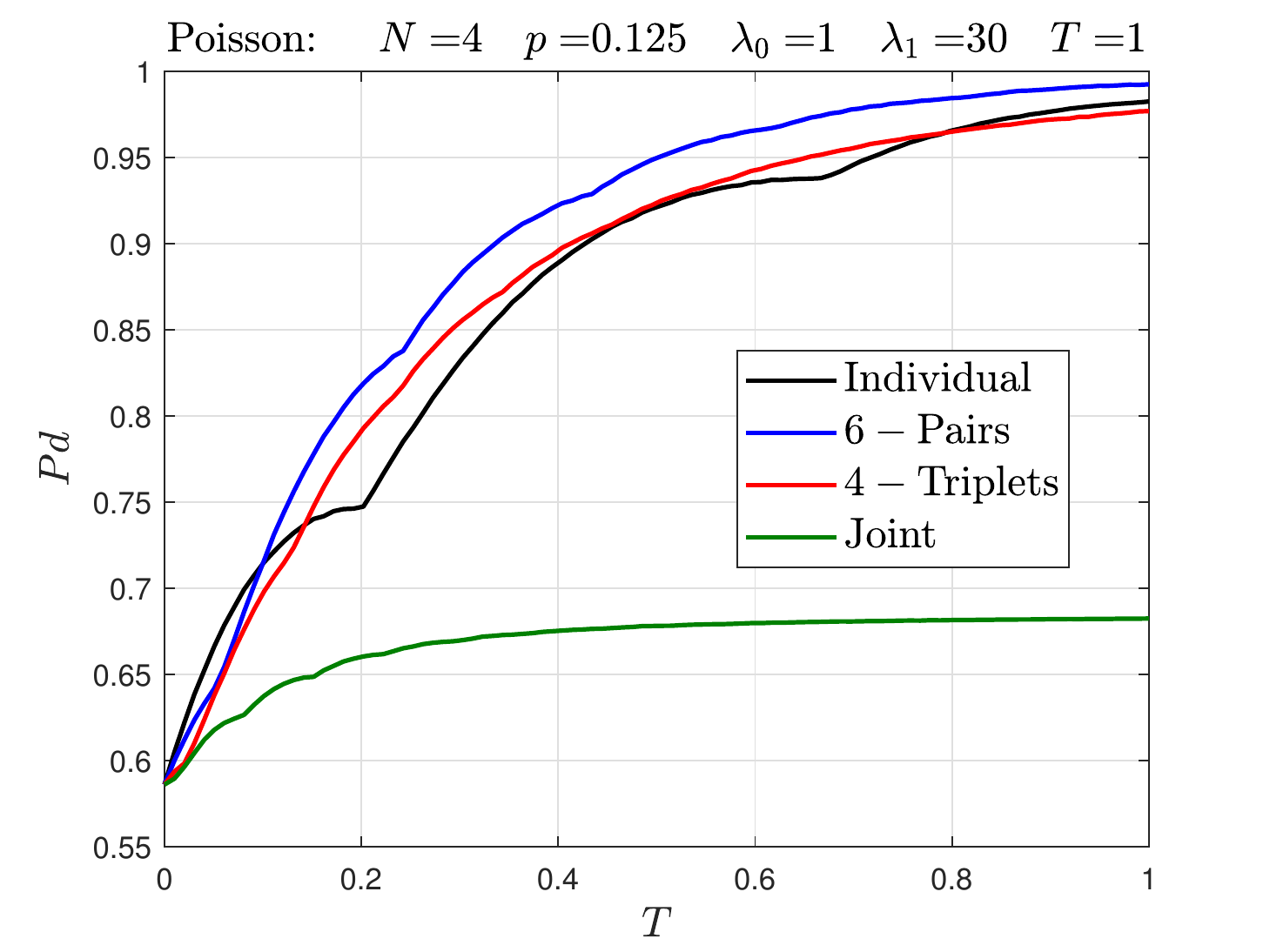}
		\caption{ }
		\label{fig6b}
	\end{subfigure} \\%
	\begin{subfigure}{.49\textwidth}
		\includegraphics[width=\linewidth]{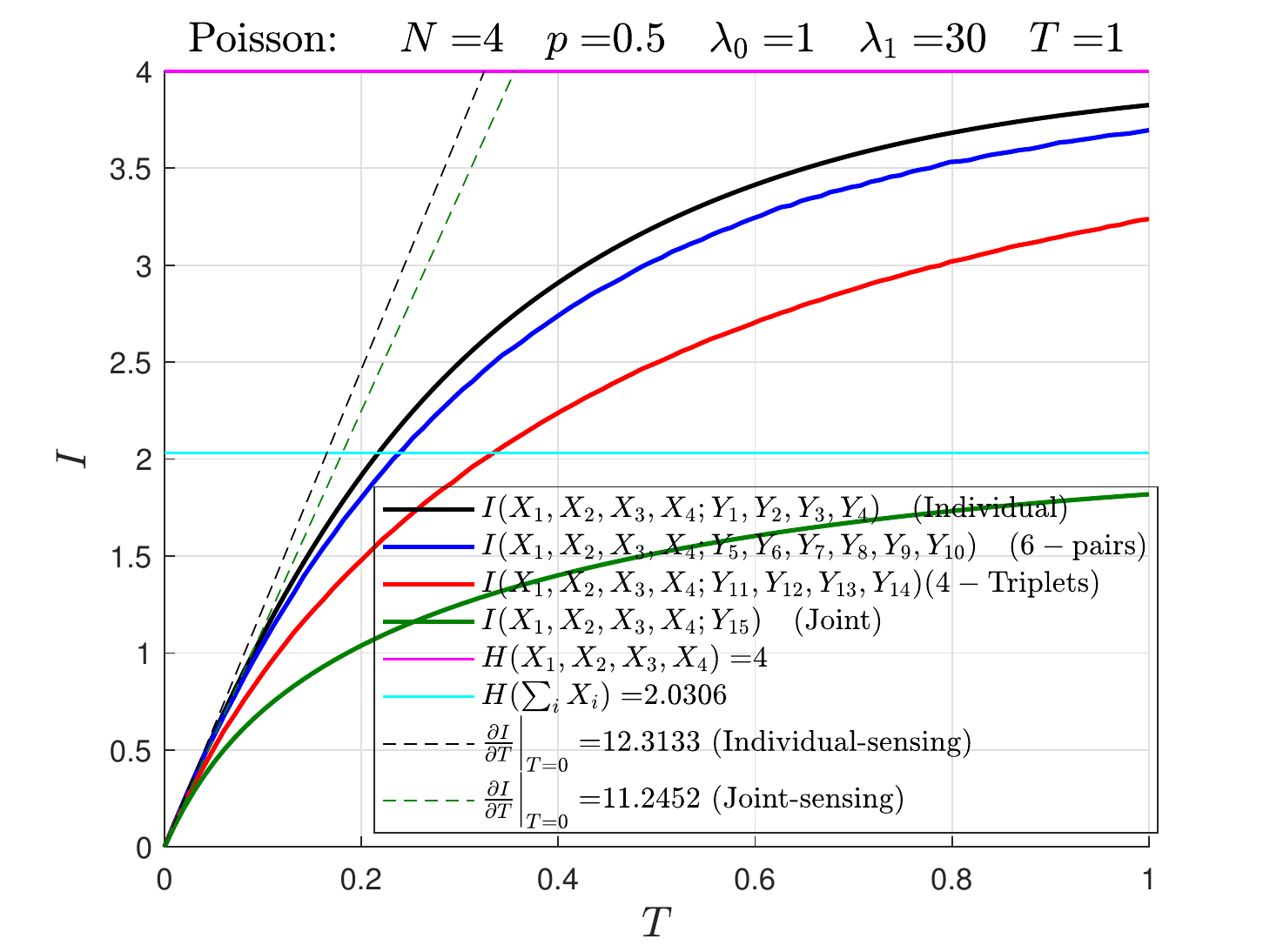}
		\caption{ }
		\label{fig6c}
	\end{subfigure} %
	\begin{subfigure}{.49\textwidth}
		\includegraphics[width=\linewidth]{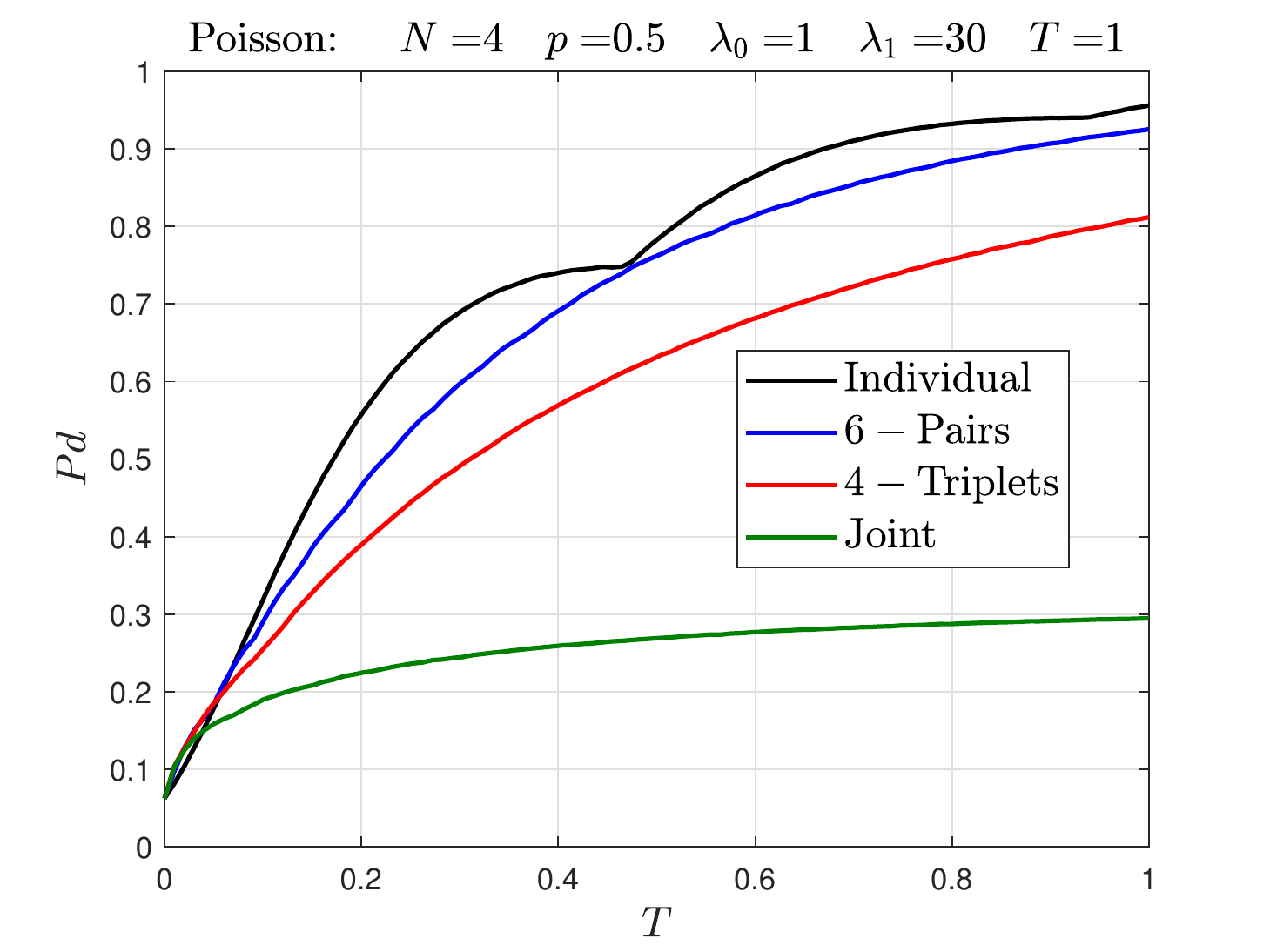}
		\caption{ }
		\label{fig6d}
	\end{subfigure}
	\begin{subfigure}{.49\textwidth}
		\includegraphics[width=\linewidth]{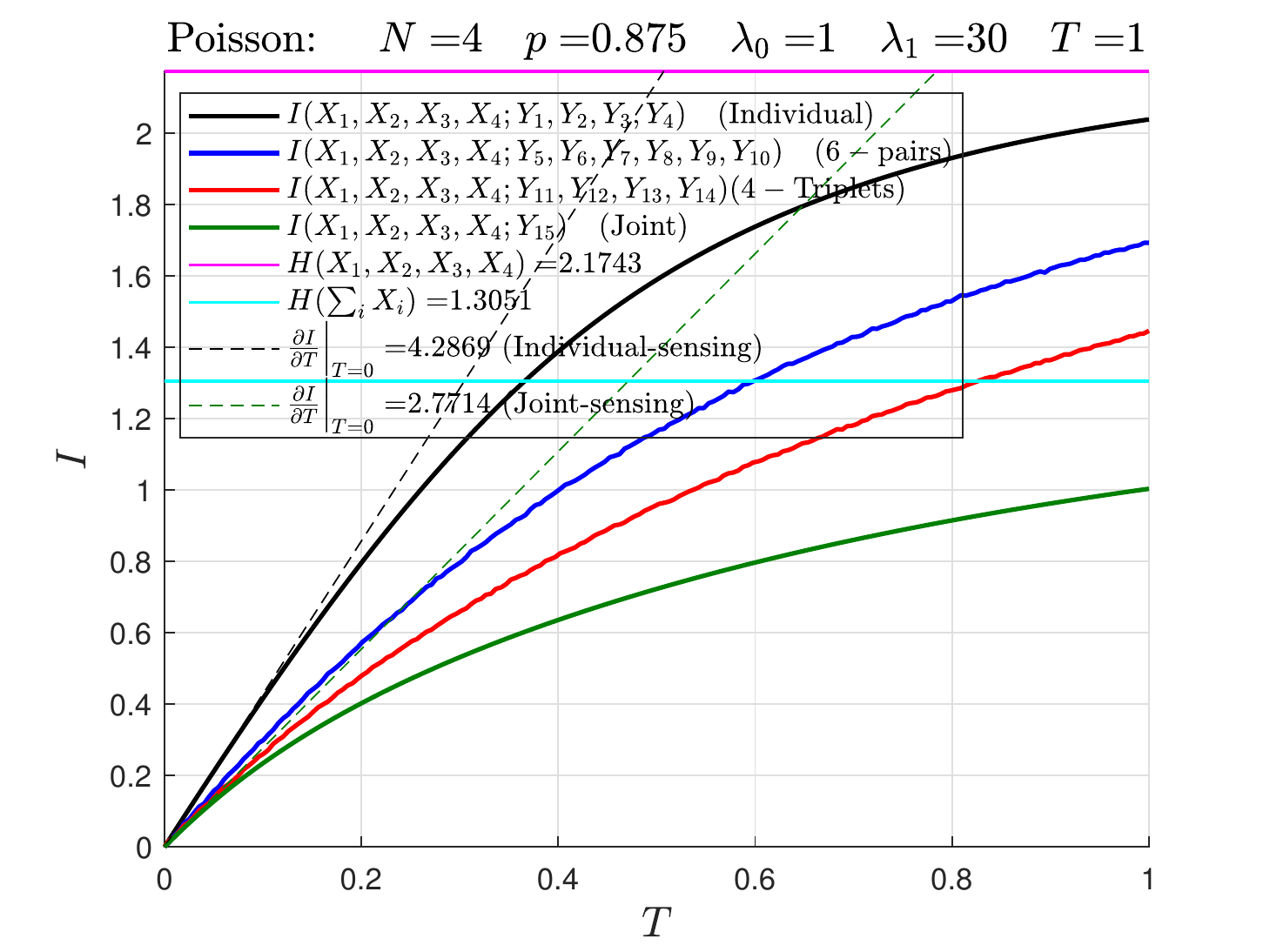}
		\caption{ }
		\label{fig6e}
	\end{subfigure} %
	\begin{subfigure}{.49\textwidth}
		\includegraphics[width=\linewidth]{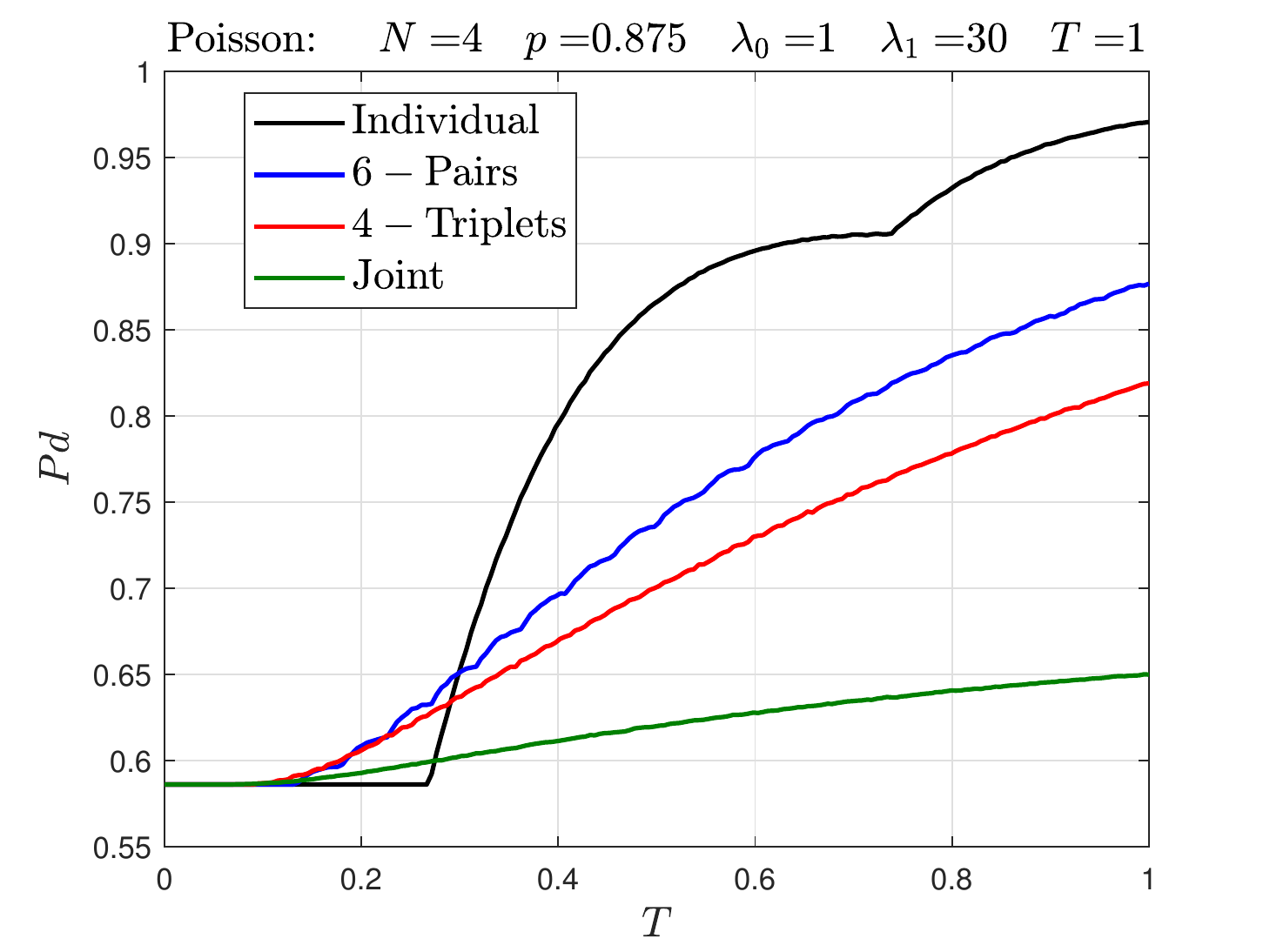}
		\caption{ }
		\label{fig6f}
	\end{subfigure} %
	\caption{Poisson channel: (Left) $I(X;Y)$ vs. $T $, (right) $P_d$ vs. $T$ for varying \emph{prior} probabilities $ p $. \todo[inline]{\texttt{\detokenize{FourTarget_Poisson_UnConstraint.m}}
			\newline   	\texttt{\detokenize{Newpoiss_nn.m}}                     }}
	\label{f6}
\end{figure}

\begin{figure*}[ht]
	\begin{subfigure}{.49\textwidth}
		\includegraphics[width=\linewidth]{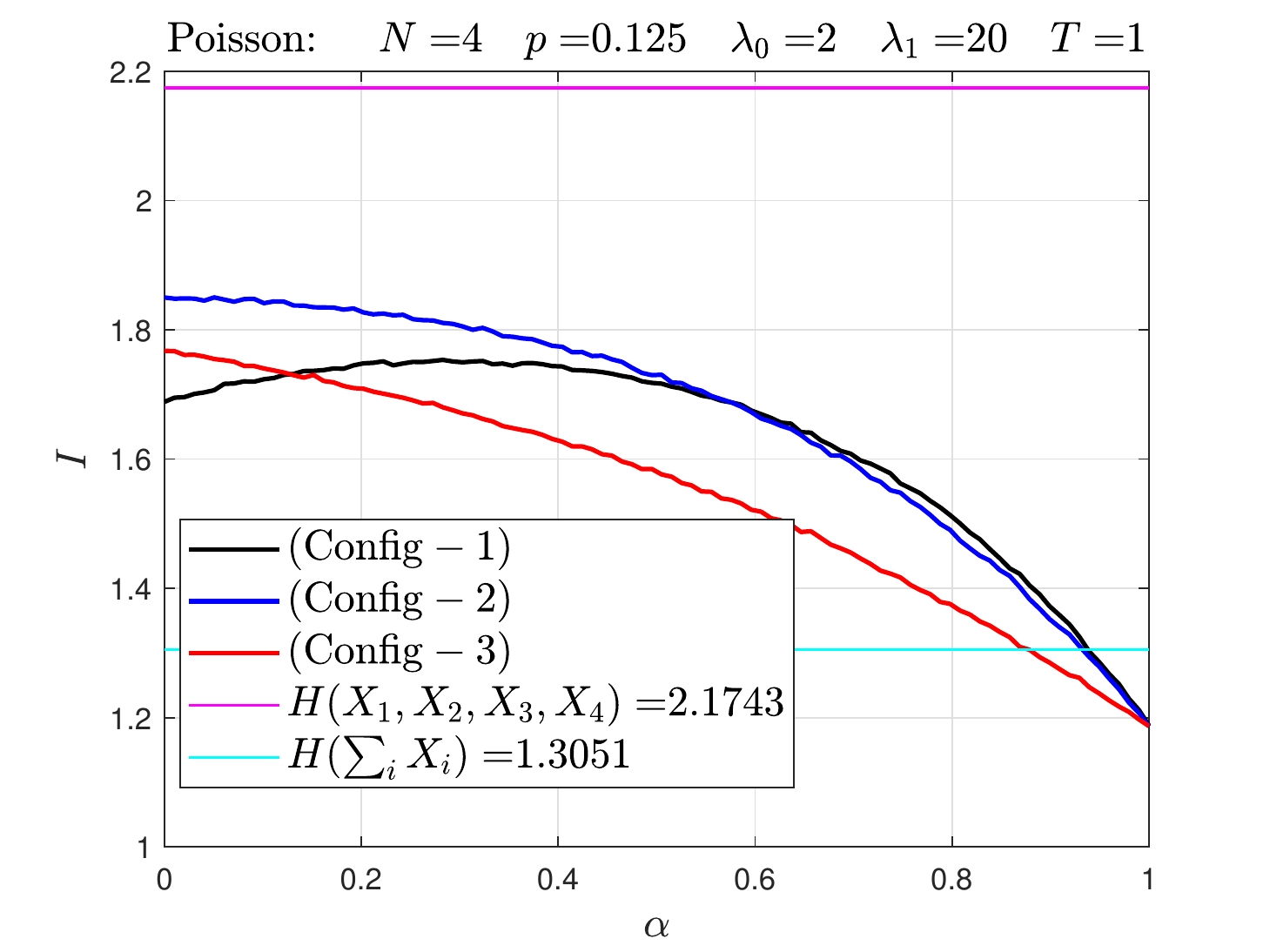}
		\caption{ }
		\label{fig1a}
	\end{subfigure} 
	\begin{subfigure}{.49\textwidth}
		\includegraphics[width=\linewidth]{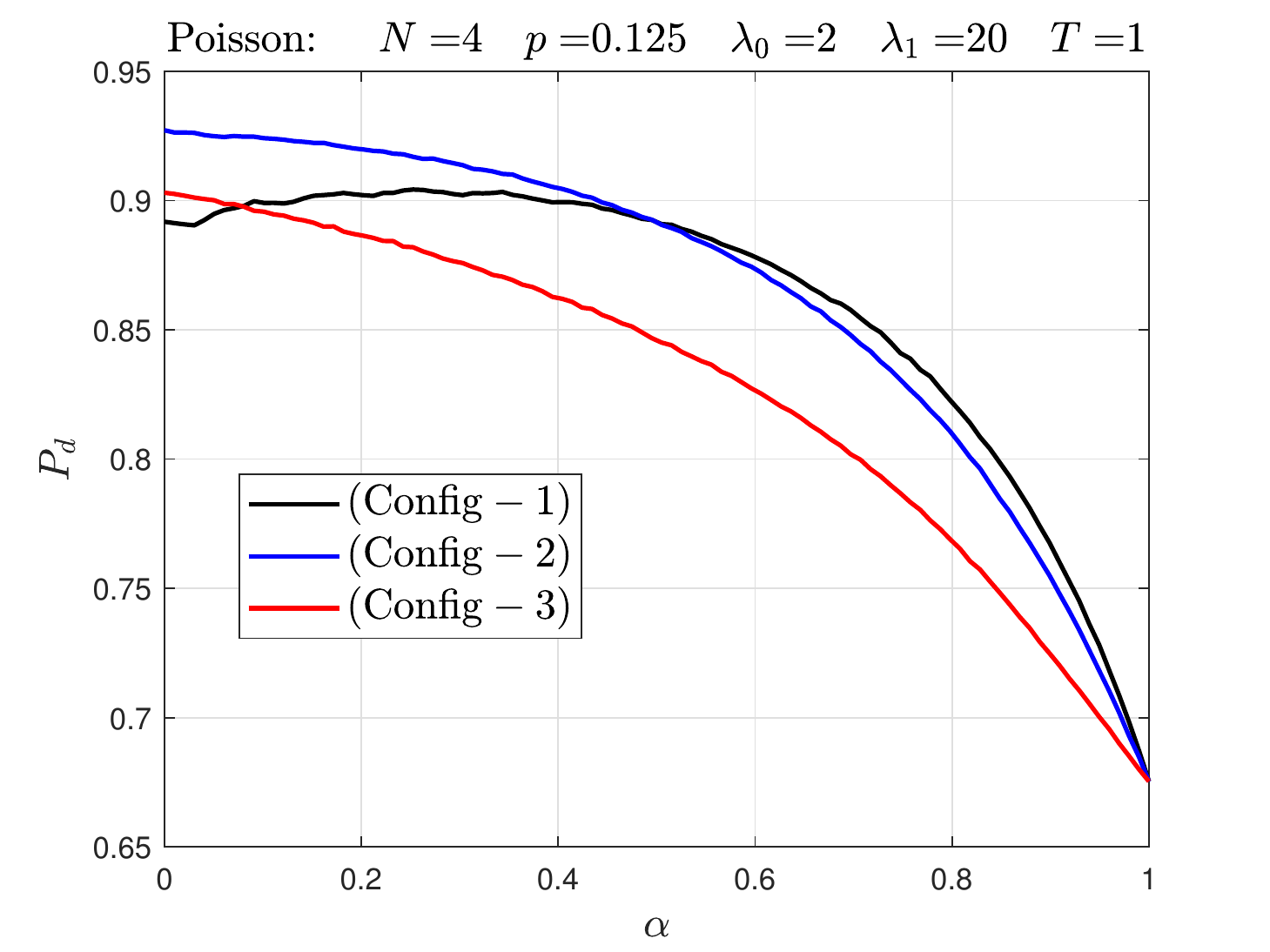}
		\caption{ }
		\label{fig1b}
	\end{subfigure} %
	\begin{subfigure}{.49\textwidth}
		\includegraphics[width=\linewidth]{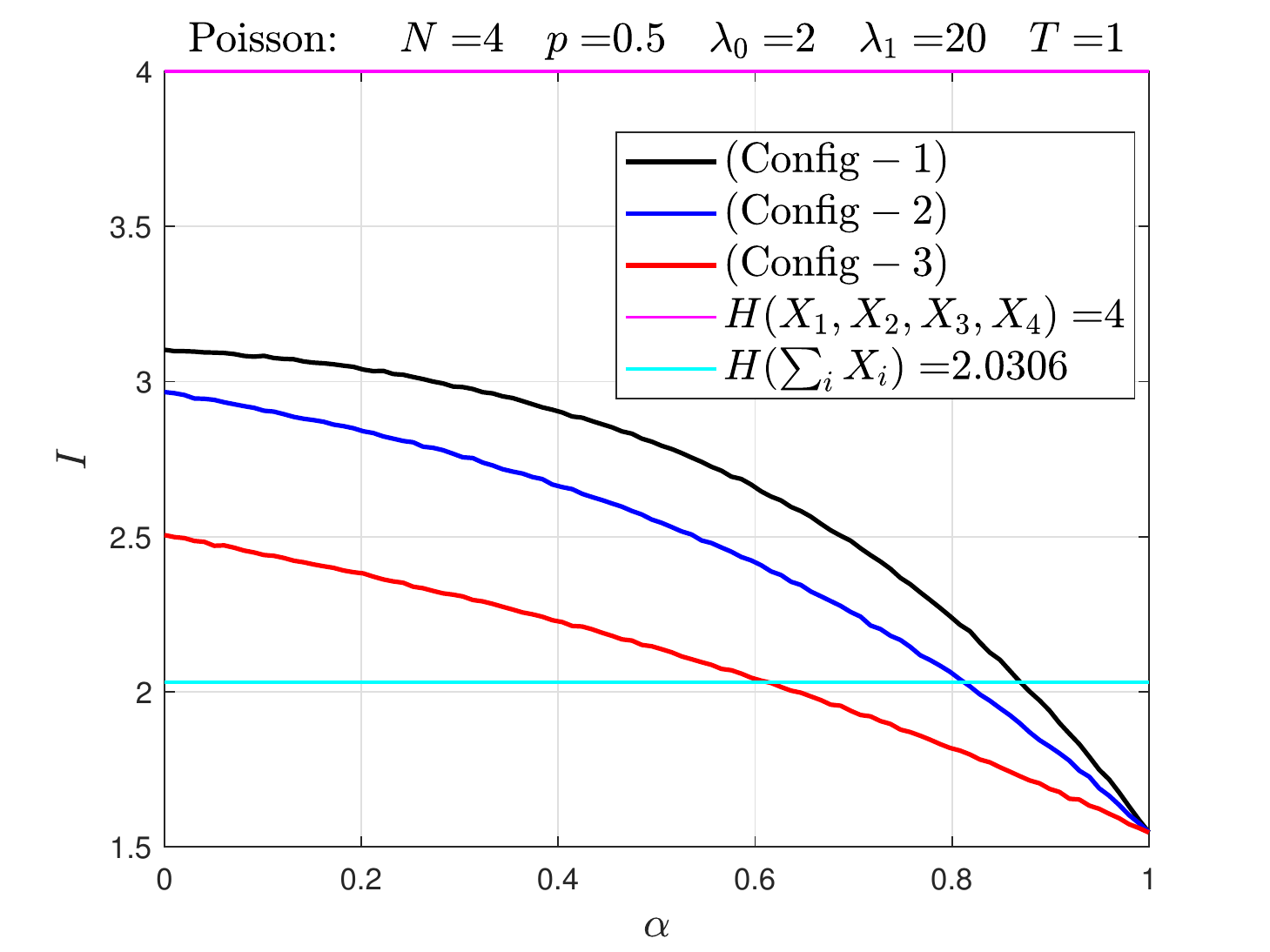}
		\caption{ }
		\label{fig1c}
	\end{subfigure} %
	\begin{subfigure}{.49\textwidth}
		\includegraphics[width=\linewidth]{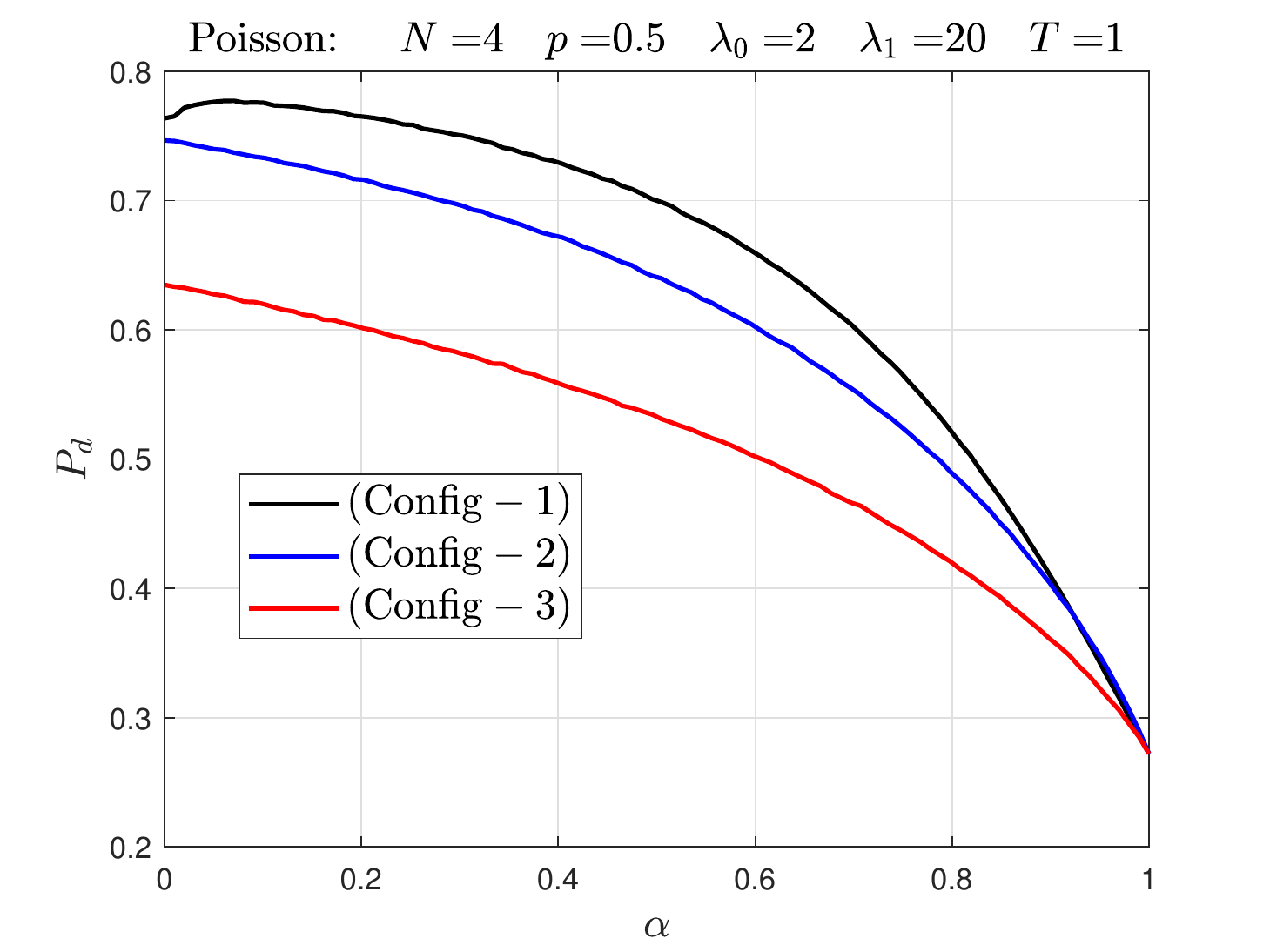}
		\caption{ }
		\label{fig1d}
	\end{subfigure}
	\begin{subfigure}{.49\textwidth}
		\includegraphics[width=\linewidth]{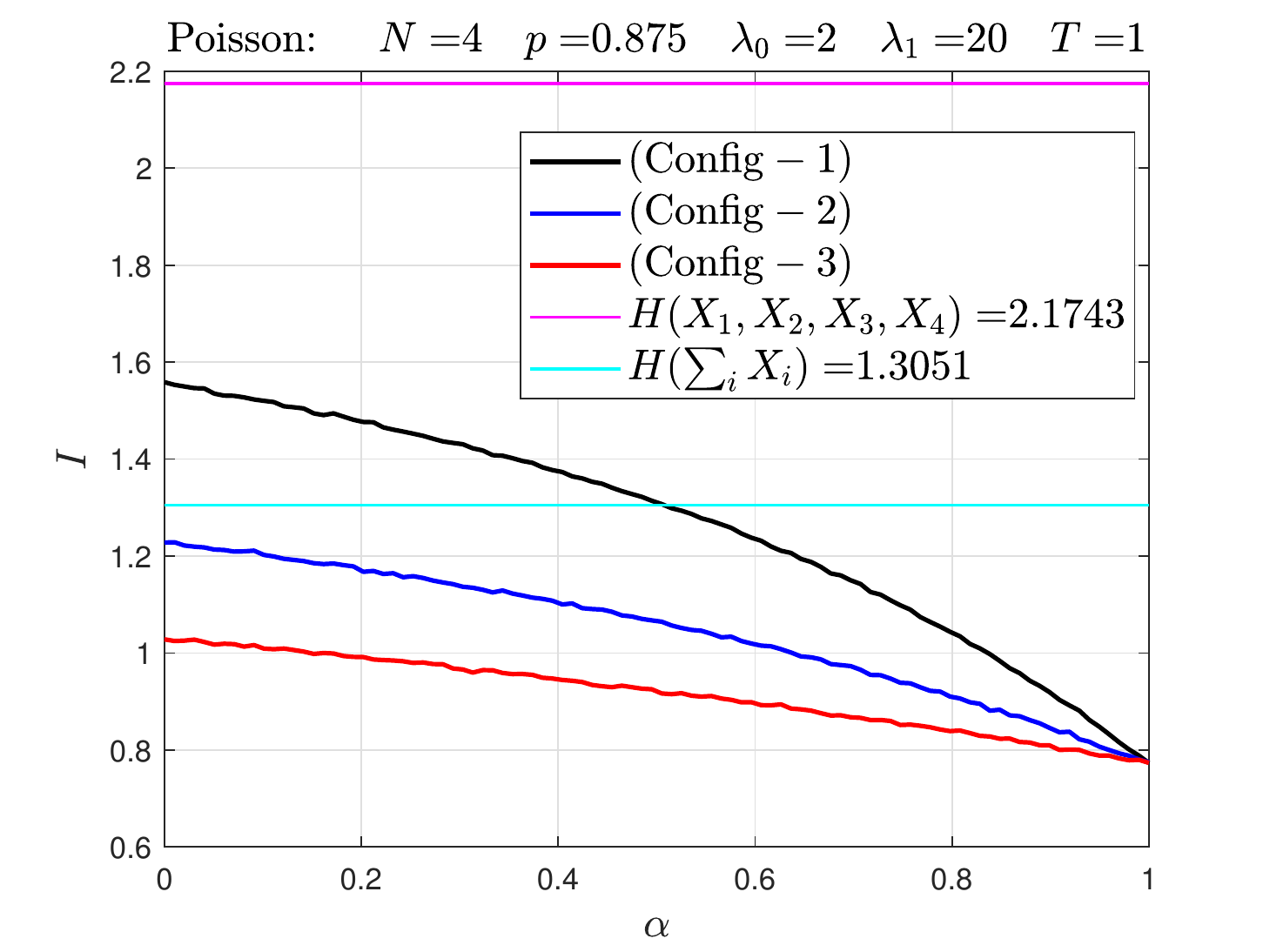}
		\caption{ }
		\label{fig1e}
	\end{subfigure} %
	\begin{subfigure}{.49\textwidth}
		\includegraphics[width=\linewidth]{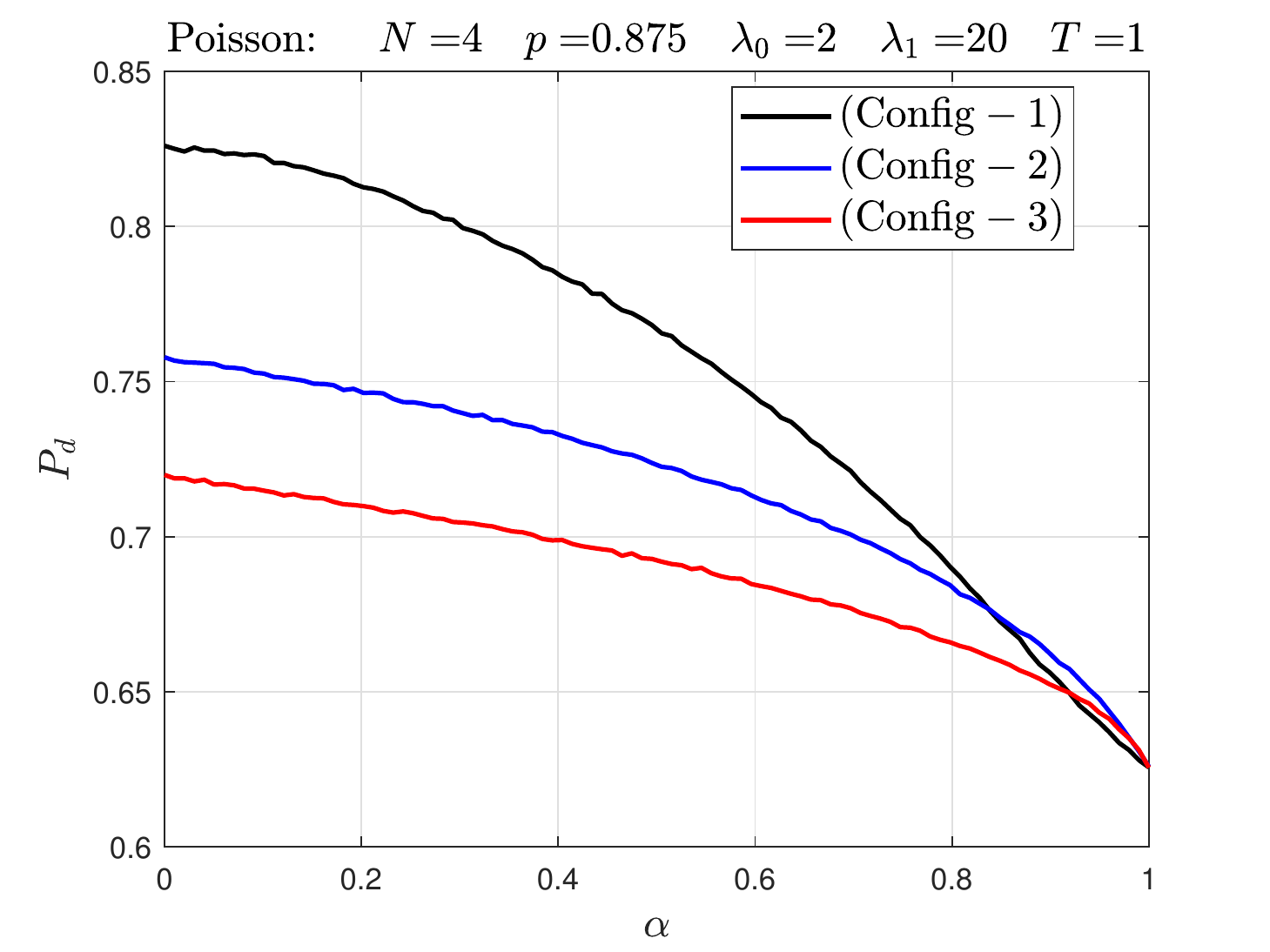}
		\caption{ }
		\label{fig1f}
	\end{subfigure} %
	\caption{\scriptsize Poisson channel : $\rm{Config-1:}  \Big(\frac{T-\alpha}{4},\frac{T-\alpha}{4},\frac{T-\alpha}{4},\frac{T-\alpha}{4},0,0,0,0,0,0,0,0,0,0,\alpha \Big) $; \\ $\rm{Config-2:}  \Big(0,0,0,0,\frac{T-\alpha}{6},\frac{T-\alpha}{6},\frac{T-\alpha}{6},\frac{T-\alpha}{6},\frac{T-\alpha}{6},\frac{T-\alpha}{6},0,0,0,0,\alpha \Big) $ and \\ $\rm{Config-3:}  \Big(0,0,0,0,0,0,0,0,0,0, \frac{T-\alpha}{4},\frac{T-\alpha}{4},\frac{T-\alpha}{4},\frac{T-\alpha}{4},\alpha \Big) $  where $0 \le \alpha \le T$ and time constraint $ \sum_{i=1}^{15} T_i=T $ for \\ $ \lambda_0=2 $, $ \lambda_1=20 $,  and varying \emph{prior} probability $ p $. \todo[inline]{\texttt{\detokenize{FourTarget_Poisson_Constraint.m}}
		\newline   	\texttt{\detokenize{Newpoiss_nn.m}}                     }}
	\label{f1}
\end{figure*}

\begin{figure*}[ht]
	\begin{subfigure}{.49\textwidth}
		\includegraphics[width=\linewidth]{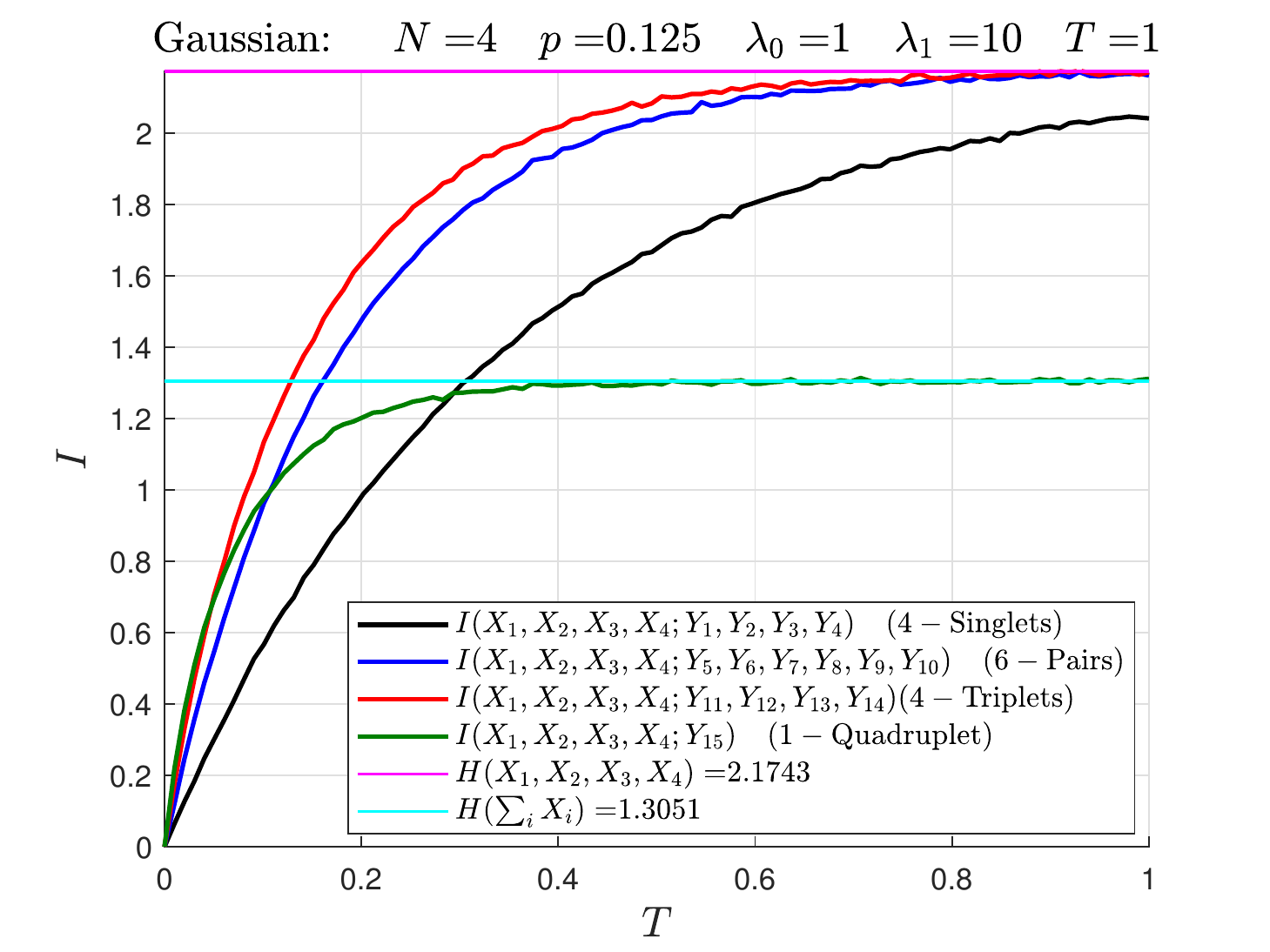}
		\caption{ }
		\label{fig7a}
	\end{subfigure} 
	\begin{subfigure}{.49\textwidth}
		\includegraphics[width=\linewidth]{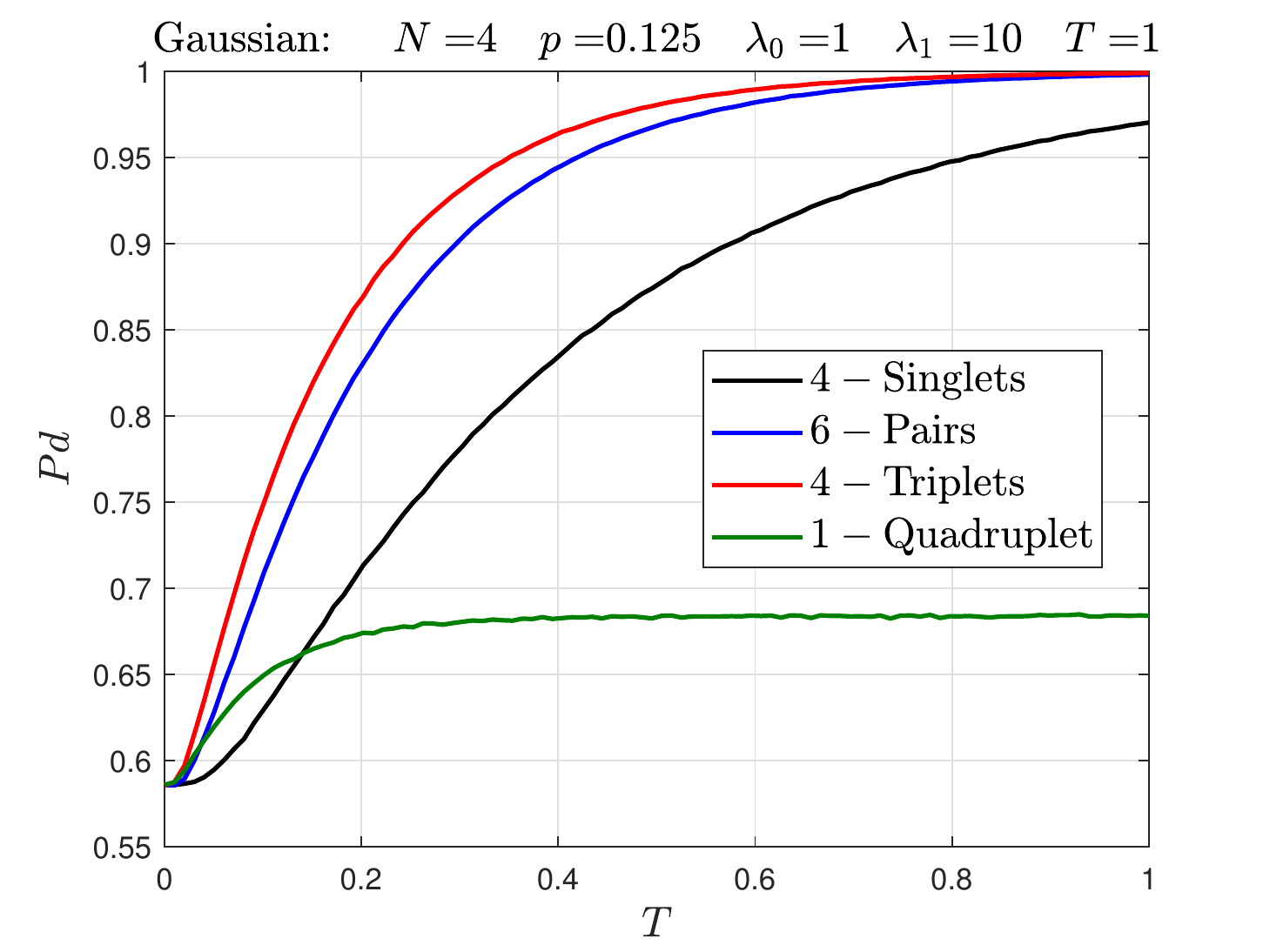}
		\caption{ }
		\label{fig7b}
	\end{subfigure} %
 	\begin{subfigure}{.49\textwidth}
 		\includegraphics[width=\linewidth]{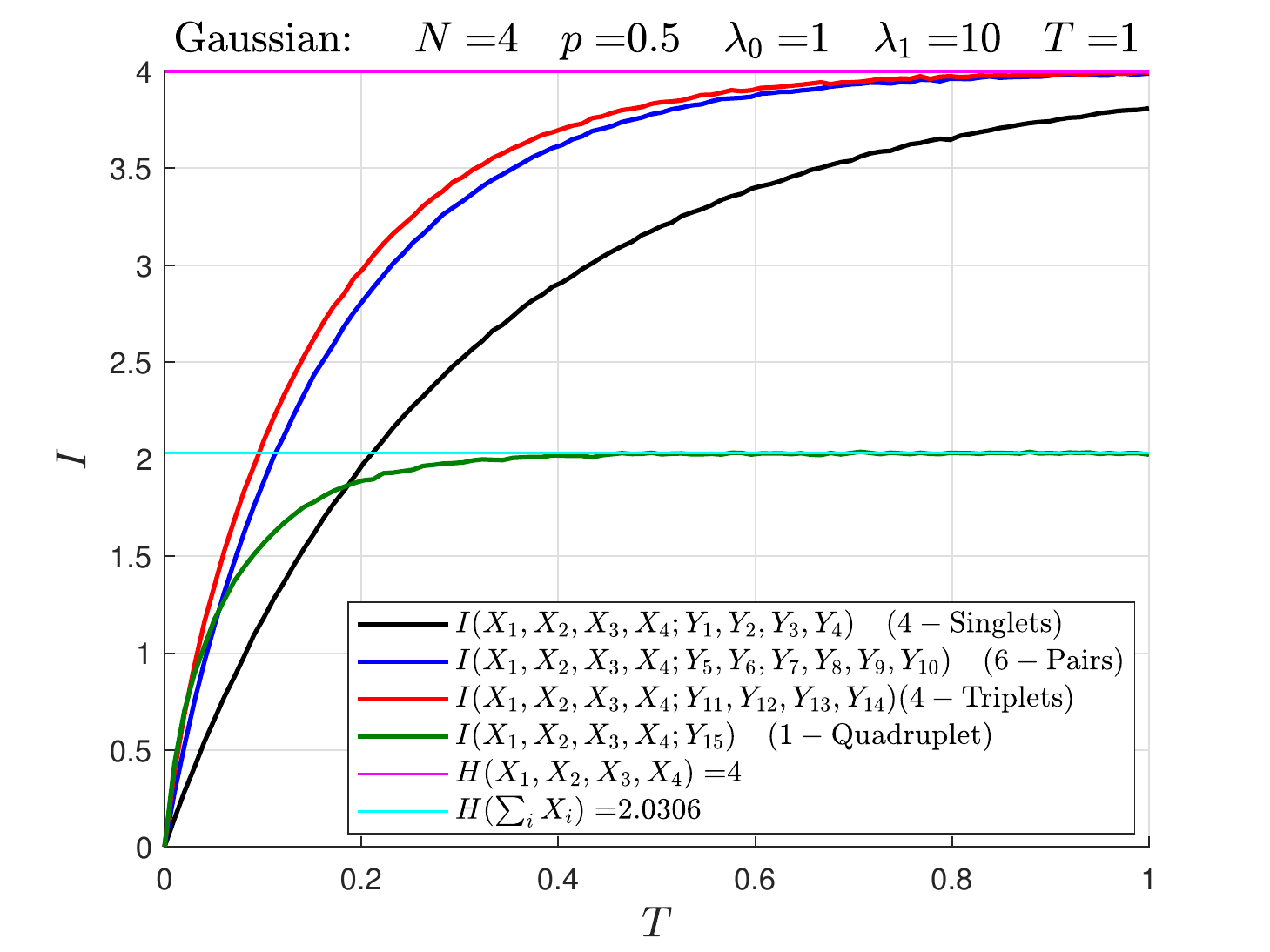}
 		\caption{ }
 		\label{fig7c}
 	\end{subfigure} %
 	\begin{subfigure}{.49\textwidth}
 		\includegraphics[width=\linewidth]{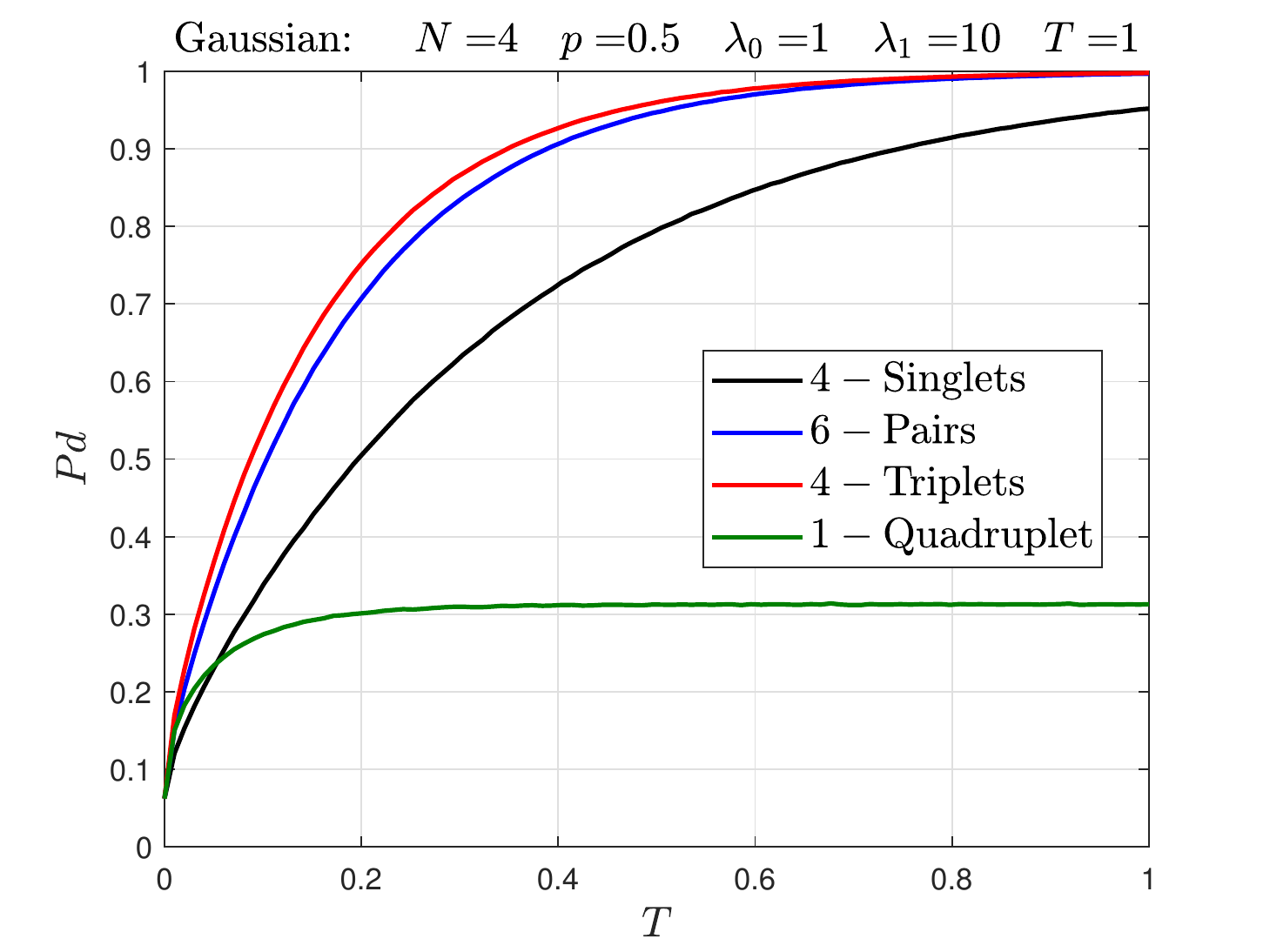}
 		\caption{ }
 		\label{fig7d}
 	\end{subfigure}
	\begin{subfigure}{.49\textwidth}
		\includegraphics[width=\linewidth]{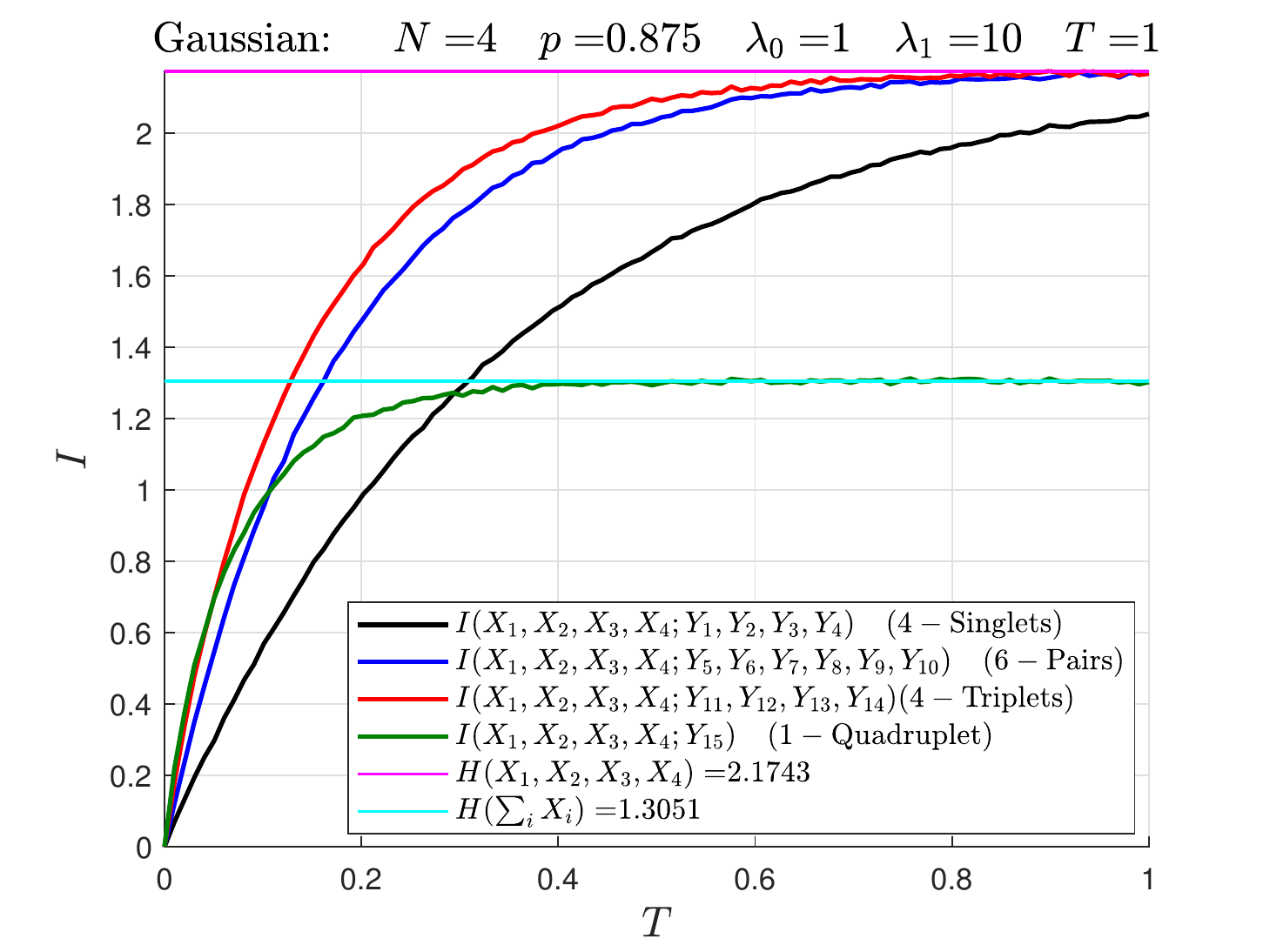}
		\caption{ }
		\label{fig7e}
	\end{subfigure} %
	\begin{subfigure}{.49\textwidth}
		\includegraphics[width=\linewidth]{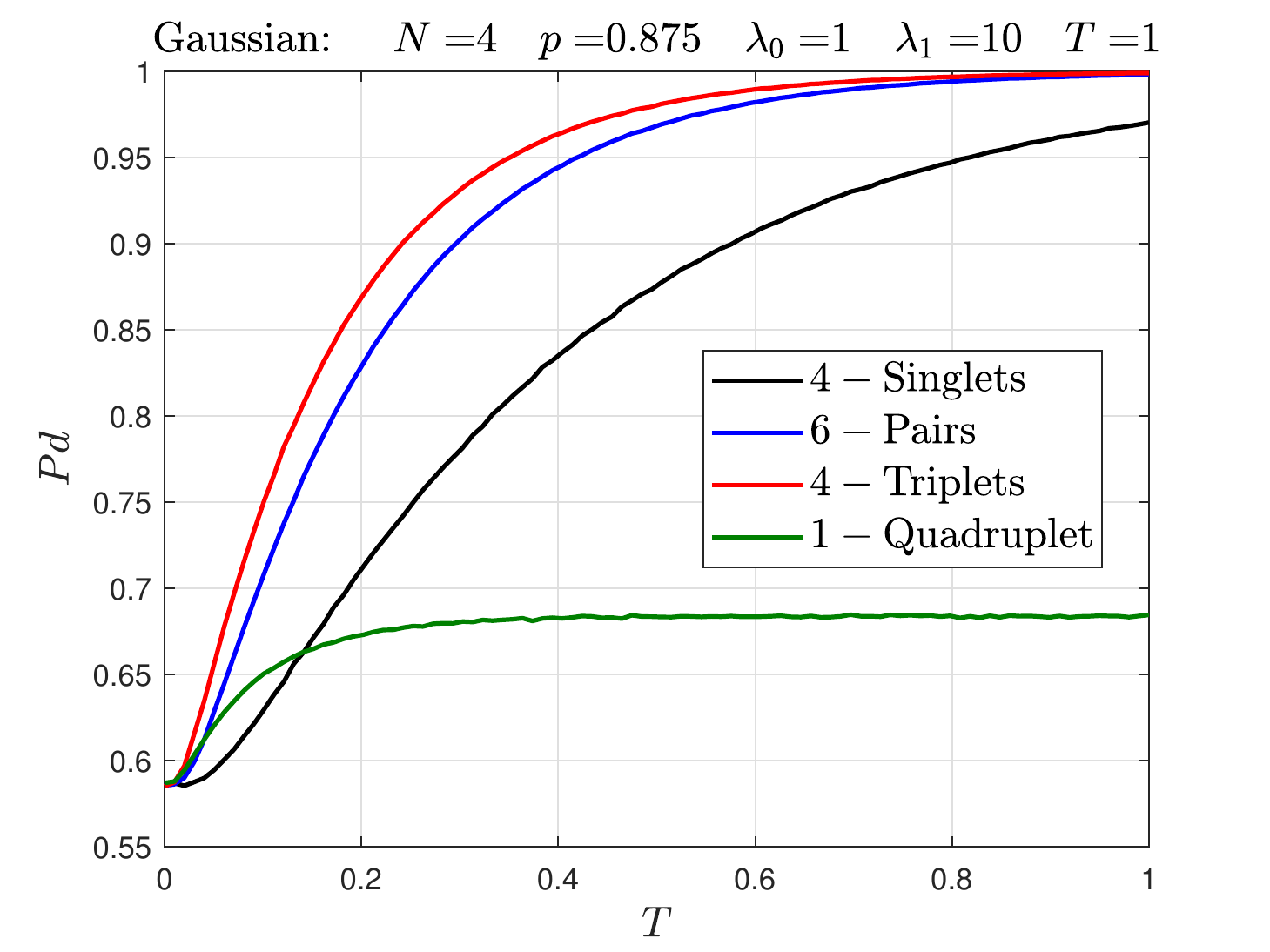}
		\caption{ }
		\label{fig7f}
	\end{subfigure} %
	\caption {Gaussian channel: (Left) $I(X;Y)$ vs. $T $, (right) $P_d$ vs. $T$ for varying \emph{prior} probabilities $ p $. \todo[inline]{\texttt{\detokenize{FourTarget_Gaussian_UnConstraint.m}}
			\newline   	\texttt{\detokenize{Newgauss_nn.m}}                     }}
	\label{f7}
\end{figure*}

\begin{figure*}[ht]
	\begin{subfigure}{.49\textwidth}
		\includegraphics[width=\linewidth]{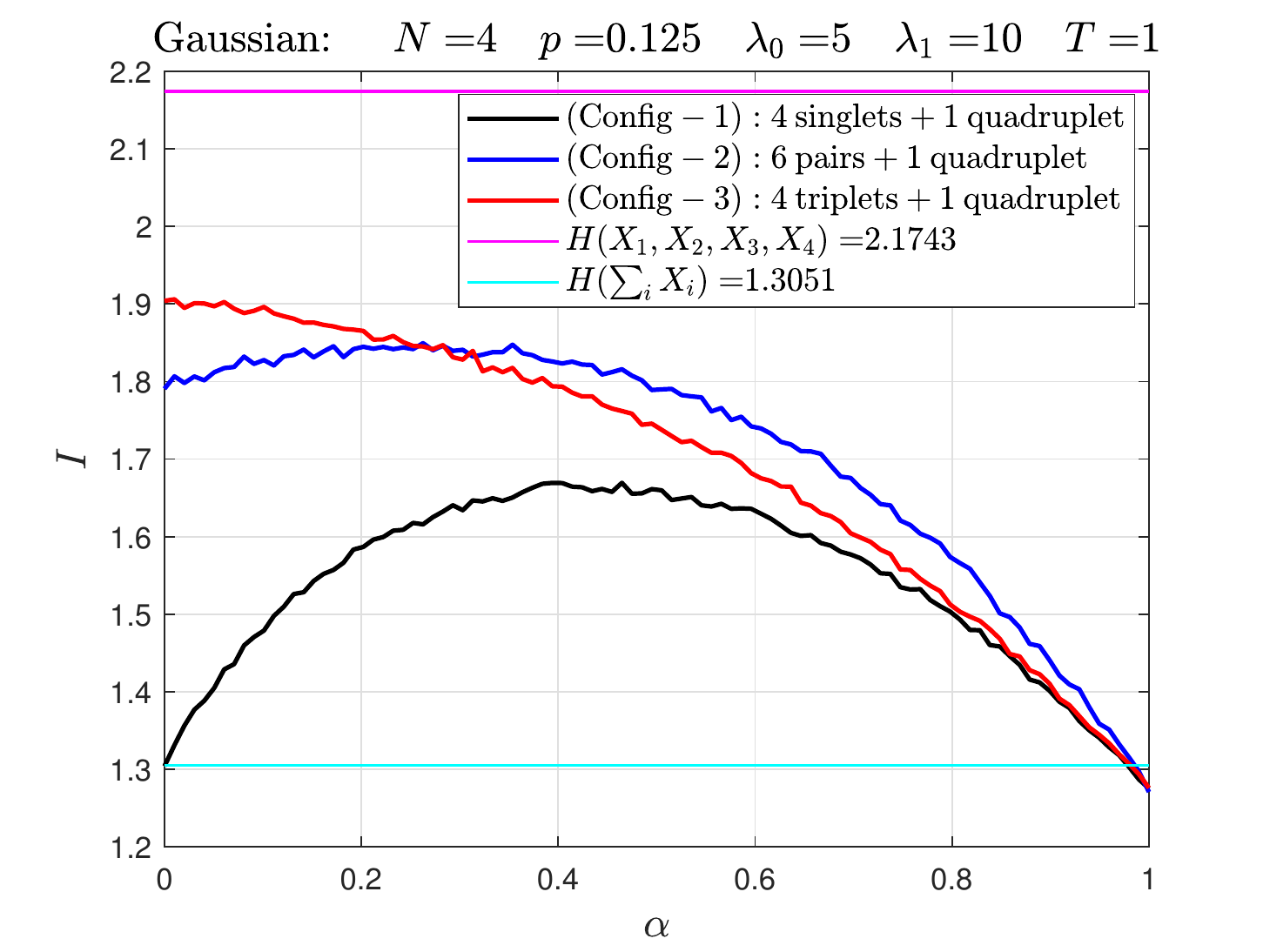}
		\caption{ }
		\label{fig2a}
	\end{subfigure} 
	\begin{subfigure}{.49\textwidth}
		\includegraphics[width=\linewidth]{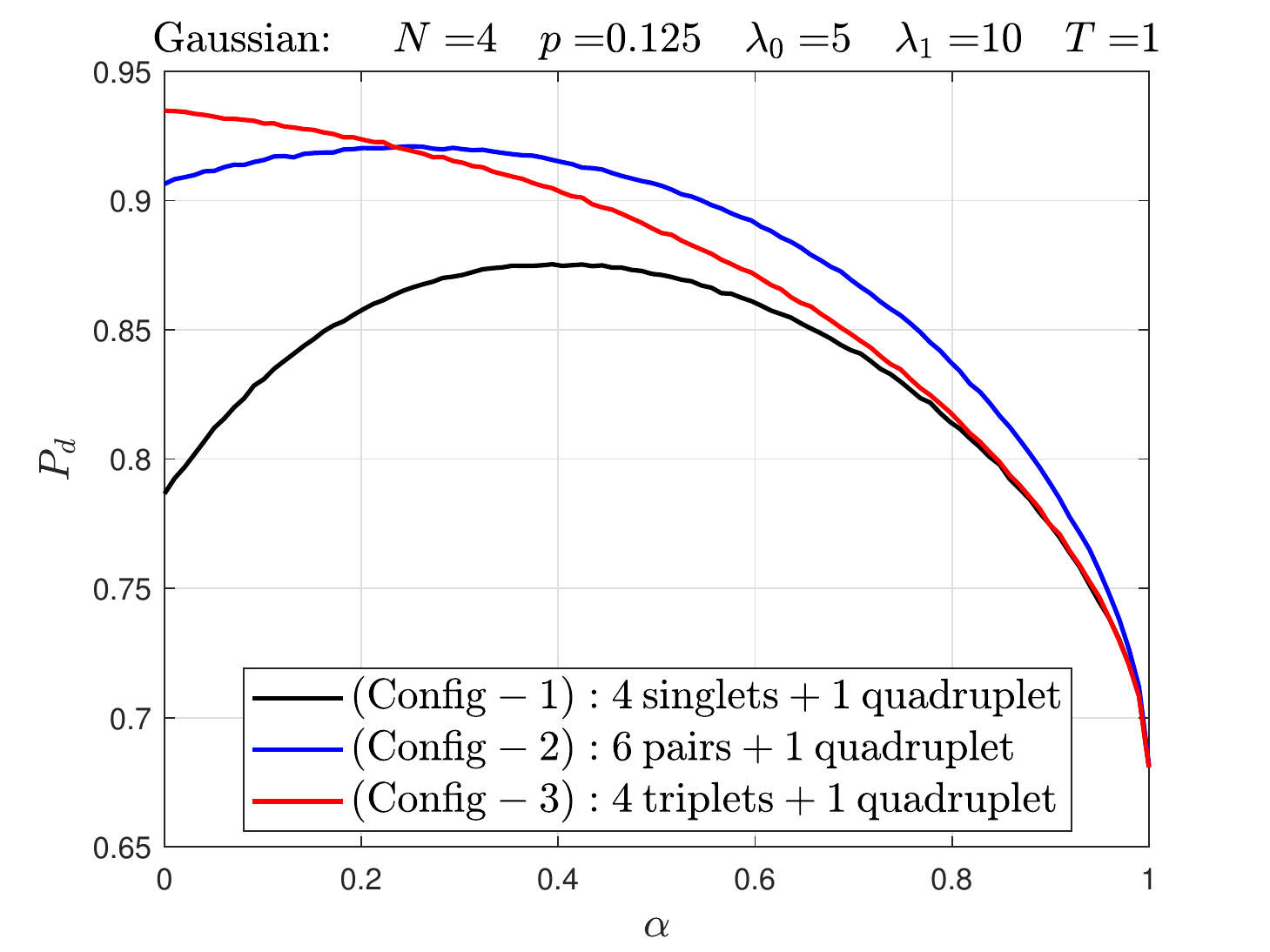}
		\caption{ }
		\label{fig2b}
	\end{subfigure} %
\begin{subfigure}{.49\textwidth}
\includegraphics[width=\linewidth]{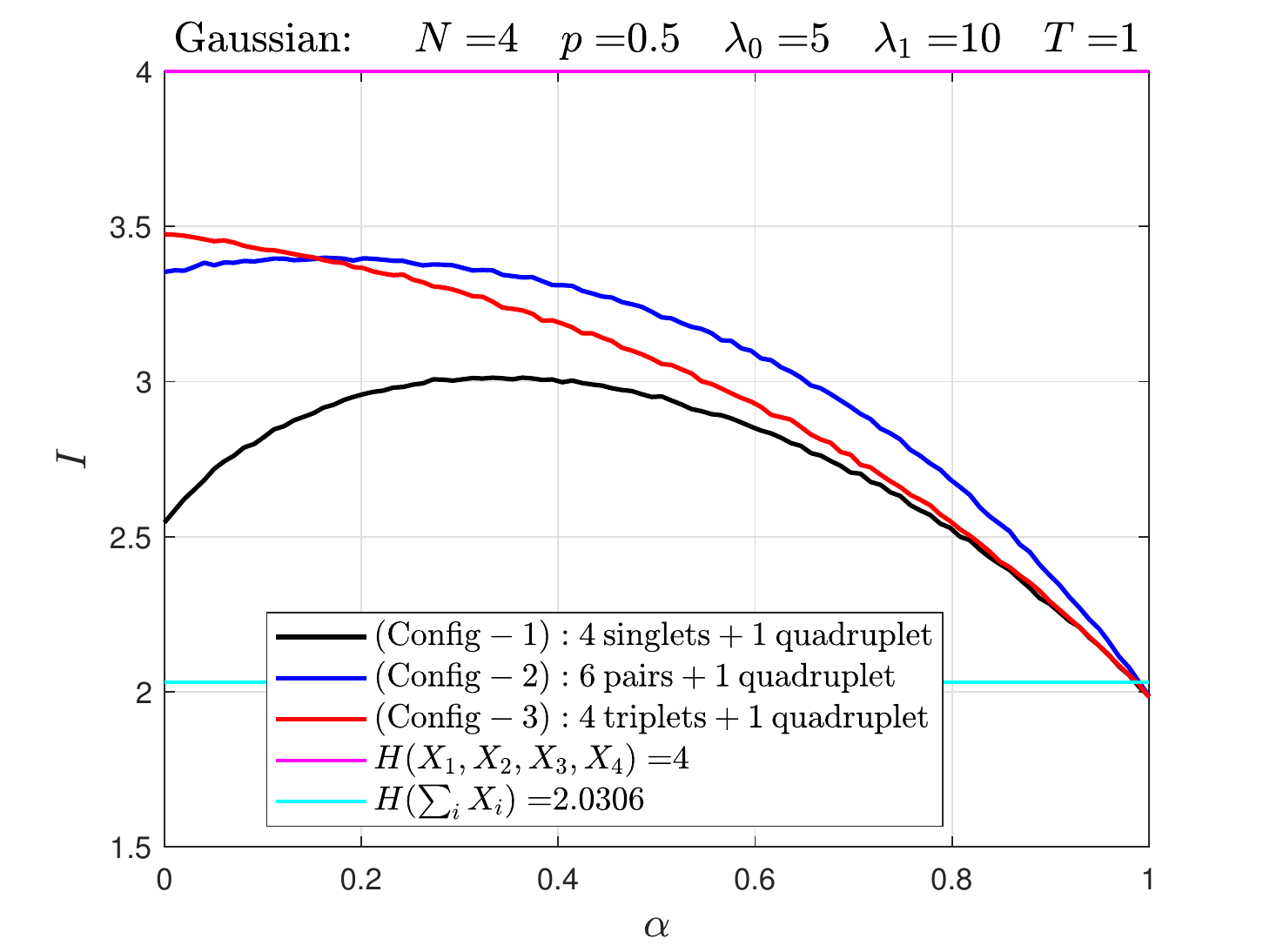}
\caption{ }
\label{fig2c}
\end{subfigure} %
\begin{subfigure}{.49\textwidth}
\includegraphics[width=\linewidth]{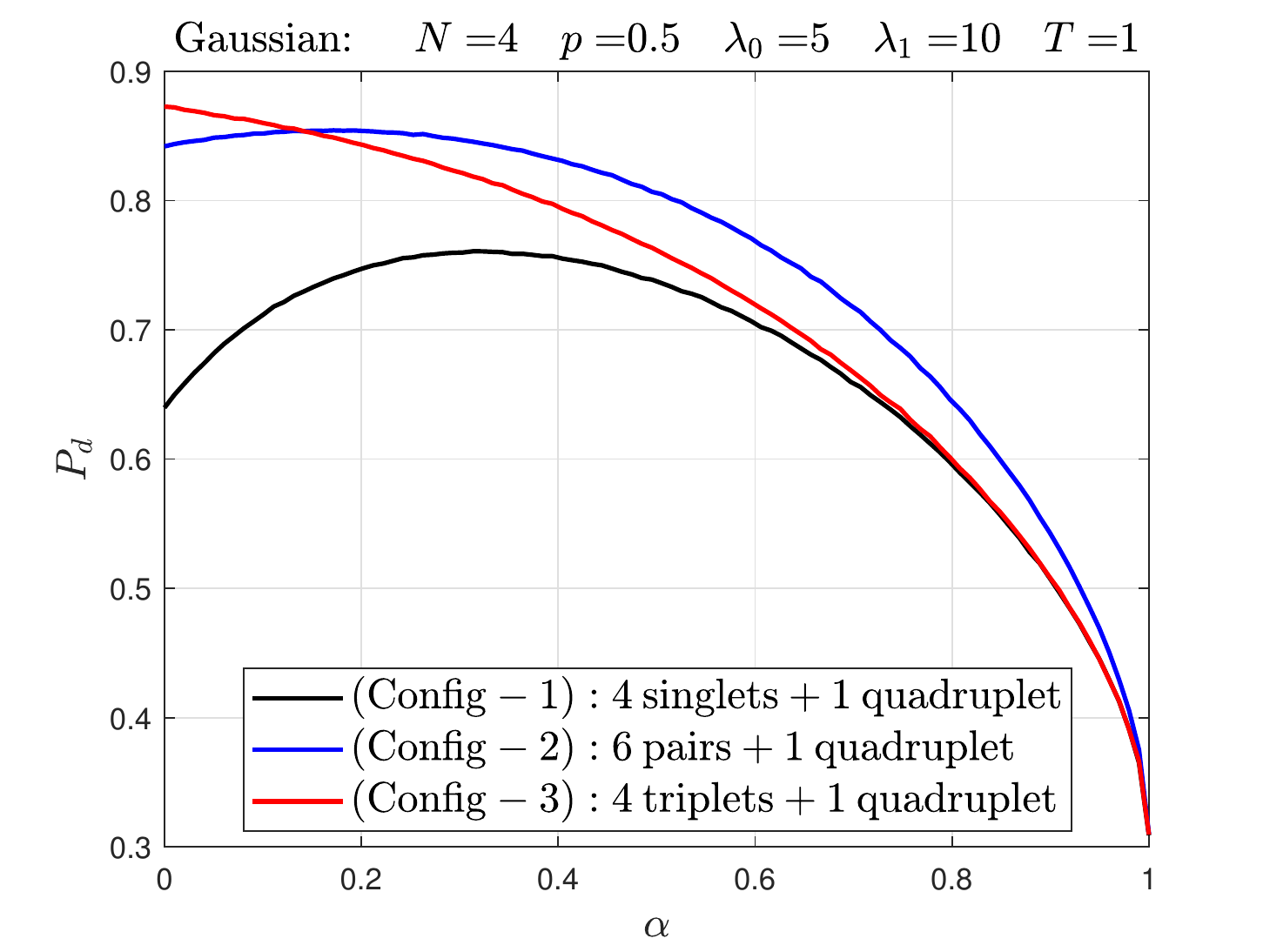}
\caption{ }
\label{fig2d}
\end{subfigure}
\begin{subfigure}{.49\textwidth}
\includegraphics[width=\linewidth]{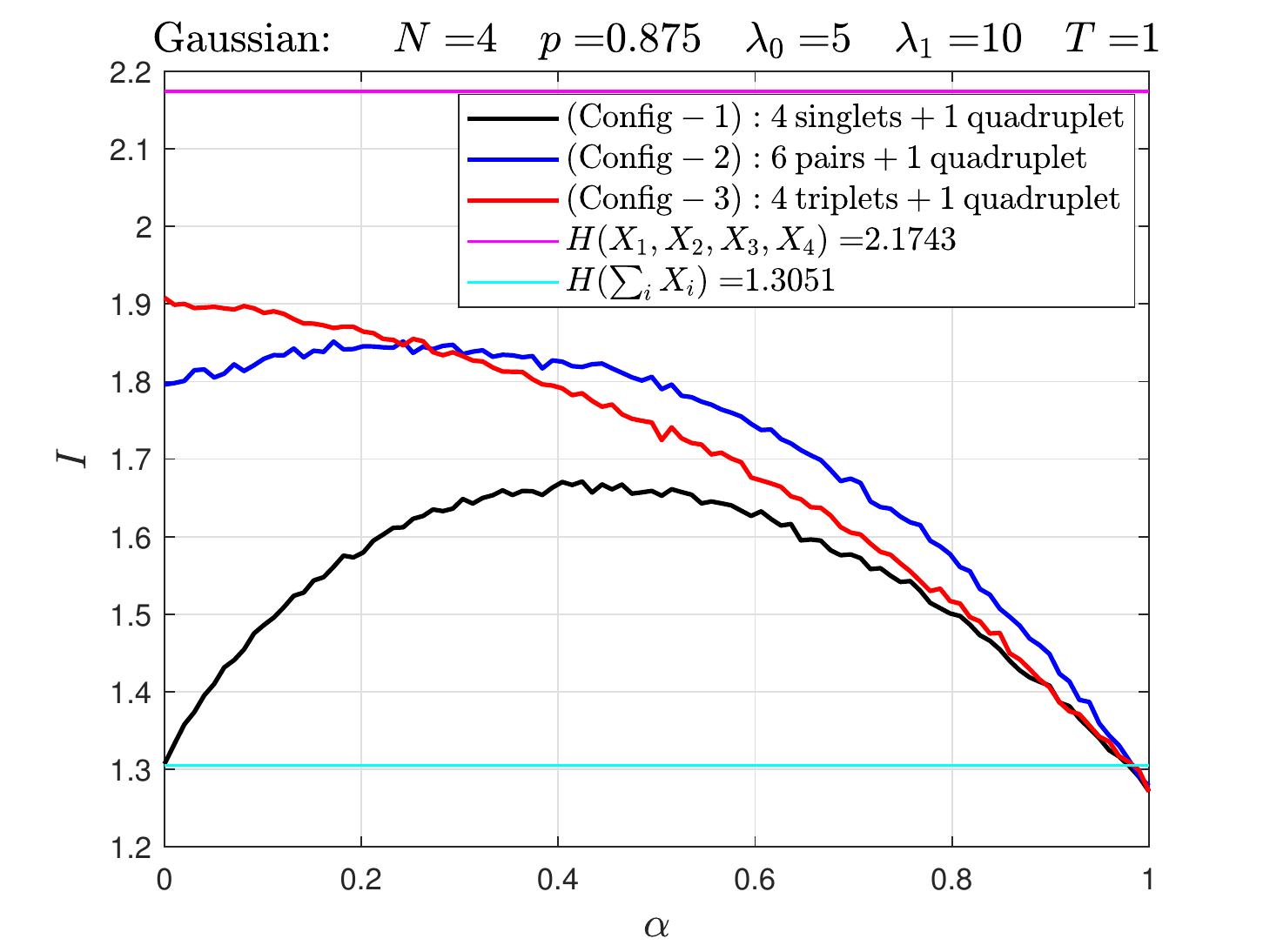}
\caption{ }
\label{fig2e}
\end{subfigure} %
\begin{subfigure}{.49\textwidth}
\includegraphics[width=\linewidth]{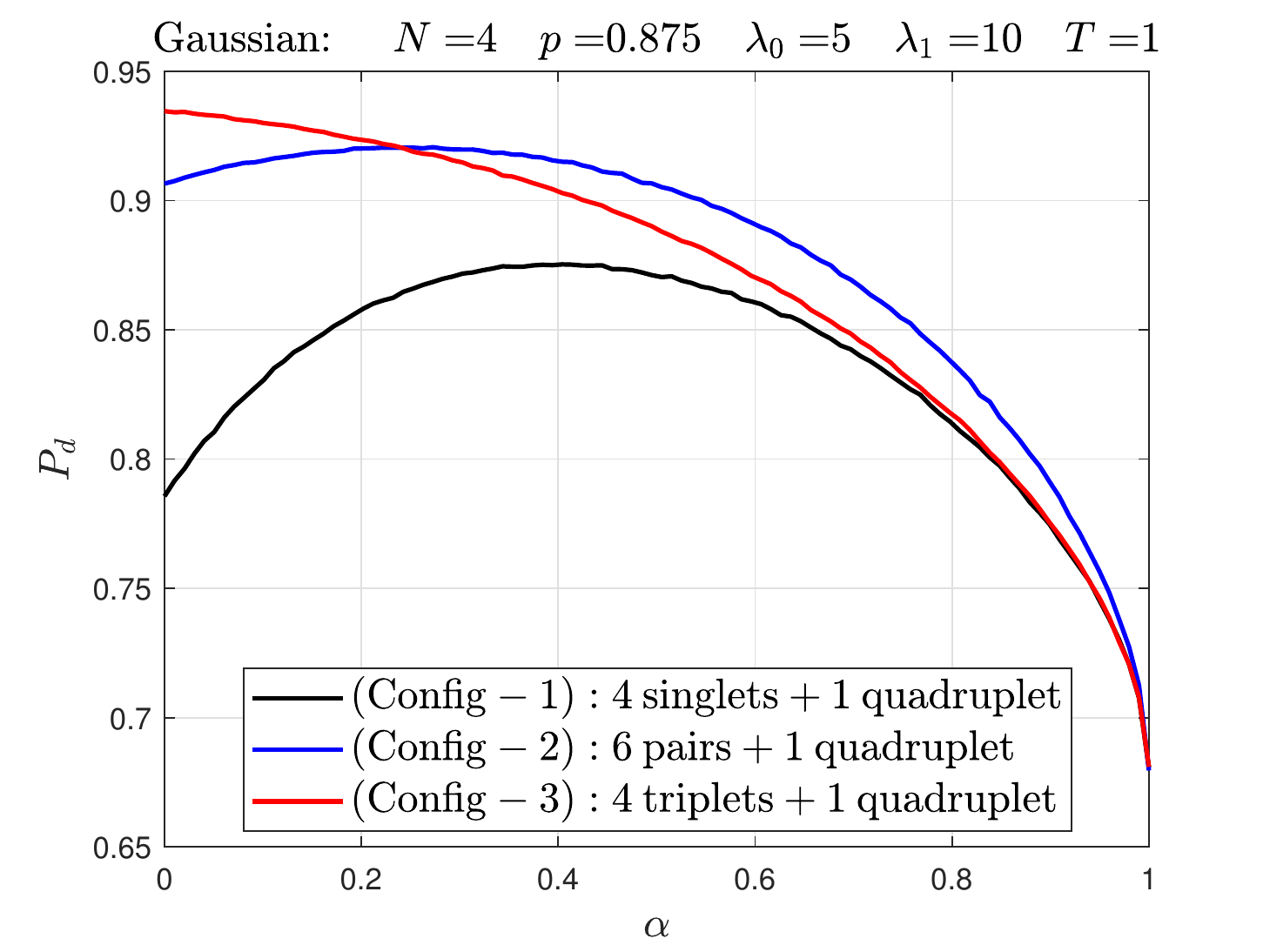}
\caption{ }
\label{fig2f}
\end{subfigure} %
	\caption{\scriptsize Gaussian channel :$\rm{Config-1:}  \Big(\frac{T-\alpha}{4},\frac{T-\alpha}{4},\frac{T-\alpha}{4},\frac{T-\alpha}{4},0,0,0,0,0,0,0,0,0,0,\alpha \Big) $; \\ $\rm{Config-2:}  \Big(0,0,0,0,\frac{T-\alpha}{6},\frac{T-\alpha}{6},\frac{T-\alpha}{6},\frac{T-\alpha}{6},\frac{T-\alpha}{6},\frac{T-\alpha}{6},0,0,0,0,\alpha \Big) $ and \\ $\rm{Config-3:}  \Big(0,0,0,0,0,0,0,0,0,0,\frac{T-\alpha}{4},\frac{T-\alpha}{4},\frac{T-\alpha}{4},\frac{T-\alpha}{4},\alpha \Big) $ where $0 \le \alpha \le T$ and time constraint $ \sum_{i=1}^{15} T_i=T $ for $ \lambda_0=5 $, $ \lambda_1=10 $,  and varying \emph{prior} probability $ p $. \todo[inline]{\texttt{\detokenize{FourTarget_Gaussian_Constraint.m}}
		\newline   	\texttt{\detokenize{Newgauss_nn.m}}                     }}
	\label{f2}
\end{figure*}

To compute the mutual information expressions for the Poisson channel given in (\ref{e21})-(\ref{e24}) and for the Gaussian channel given in (\ref{e41})-(\ref{e44}), we have utilized a Monte Carlo method. For any of the time settings, under any sensing scheme, we first generate $10^{6}$ samples from the respective Poisson mixture pmf (or Gaussian mixture pdf). These samples are generated in a manner that based on the prior of each component Poisson multivariate (or component Gaussian multivariate), we took the same percent of samples from that component. Further, as in each component the random variables are mutually independent, this simplifies the samples' generation from any component. After the samples are generated from any component, for fixed time-proportions $(T_1,T_2, \cdots T_{15}) $ and given model parameters ($\lambda_0, \lambda_1, T, p$), we calculated the $P(Y)$ as given in (\ref{AL80}) and (\ref{AL8}). From the computational-time point-of-view, this calculation of $P(Y)$ is most time-consuming than any other step and this is due to the calculation of sixteen $15-$dimensional multivariate components involved in mixture distribution functions. $\operatorname{Log}_2$ is then taken of the $10^6$ points of $P(Y)$ before taking the average as given in (\ref{AL4}). Once we calculated the $H(Y)$, then comes the conditional entropy $H(Y|X)$. For the conditional Gaussian entropy the expression is simple as given in (\ref{AL9}). For the conditional Poisson entropy we first truncate the conditional Poisson pmfs of each variable $Y_i$ to a sufficiently large value which is calculated as $ y_{i_{max}}=2 \cdot \operatorname{PoissCDF}^{-1}(1-1.110223024625157 \cdot 10^{-16},\lambda)$, where  $ y_{i_{max}} $ is the truncation point and $\operatorname{PoissCDF}^{-1}(m,\lambda)$ is the inverse Poisson cumulative distribution function (cdf) with parameter $\lambda$ and at point $m$. After the truncation; a finite summation for the individual variable $y_i$ can be calculated easily i-e $H(Y_i|\lambda)= -\sum_{j=0}^{y_{i_{max}}}  \operatorname{Log}_2 [\operatorname{Poiss}(j;\lambda)] \cdot \operatorname{Poiss}(j;\lambda)$. Conditional entropy for Poisson channel can then be readily calculated from (\ref{AL90}).
For the MAP detection, we use the same samples, for posterior probabilities of each of the $15-$hypothesis, that are previously used for the calculation of mutual information. We had generated the samples from each of the component with the same proportion as defined by the prior of that component and then calculated the joint probability of that sample point $Y_s$ with each hypothesis. Deciding in favor of the hypothesis for which the maximum of the joint probability $P(Y_s,\mathcal{H}_i)$ happens among $16$ such probabilities; as given in (\ref{e11}).
The computed results for mutual information and Bayes probability of total correct detections are shown in fig. (\ref{f6}), (\ref{f1}), (\ref{f7}), and (\ref{f2}).

  \section{Conclusion} \label{con}
   In this work, a sensor scheduling problem for four target detection in a vector Poisson and Gaussian channel was considered using metrics of mutual information $I$ and Bayes risk with $0-1$ cost.
   
   First, four sensing schemes: individual-sensing; pairs-sensing; triplets-sensing; and joint-sensing were considered with the total given time $T$ being variable. It was shown that mutual information between input and output is concave w.r.t given time, (and irrespective of any other model parameters) for either of the two channels. It is further noted that for the Poisson channel; individual sensing is the best among the four strategies if prior $p\ge 0.5$ from $I$ perspective. However, in the equivalent Bayesian risk minimization problem, neither the concavity of Bayesian probability, $P_d,$ of total correct detections w.r.t time is observed nor individual sensing is always found to be the best among others. Whereas for the Gaussian channel it is the triplets-sensing scheme that almost outperform any other sensing-scheme and Bayesian $P_d$ is not consistent with the $I$ computational results.
    
    Secondly, in another constrained configuration: where total time $T$ is always held fixed while linearly distributed between joint sensing:  and individual sensing; pairs-sensing and triplets-sensing. From computations; concavity of $I$ is observed w.r.t time shifting parameter $\alpha$.  For the Poisson problem from the $I$ perspective; it is again the individual-sensing that outperforms any other configuration for $p \ge 0.5$. This is not very much consistent from the Bayes detection perspective, however. Pair-wise sensing is more beneficial than individual sensing for prior close to zero. In the Gaussian channel it is the triplets-sensing scheme that is the best among any other scheme and this is evident from both $I$ and $P_d$ metrics. It is shown that $I$ is concave in $\alpha$.
    
The authors are interested in knowing why time-divisions: $T_1=T_2=T_3=T_4$, and $T_5=T_6=T_7=T_8=T_9=T_{10}$ and $T_{11}=T_{12}=T_{13}=T_{14}$ are better than being not equal in respective groups?  
\bibliographystyle{IEEEtran}
\bibliography{ReferencesGlobal} 

\end{document}